\documentclass[manuscript,screen,nonacm]{acmart} 
\usepackage{xcolor}
\setcopyright{none}
\usepackage[nolist]{acronym}
\begin{acronym}
    \acro{HCI}{Human-Computer Interaction}
    \acro{CSCW}{Computer-Supported Collaborative Work}
    \acro{CS}{Computer Science}
    \acro{DS}{Data Science}
    \acro{LLM}{Large Language Model}
    \acro{ML}{Machine Learning}
    \acro{EDA}{Exploratory Data Analysis}
    \acro{HMM}{Hidden Markov Model}
    \acro{AST}{Abstract Syntax Tree}
    \acro{SVM}{support vector machine}
\end{acronym}

\usepackage{dirtytalk}
\usepackage{indentfirst}
\usepackage[labelformat=simple, labelsep=colon]{subcaption}

\usepackage{geometry}
\usepackage{graphicx}
\usepackage{array}
\usepackage{booktabs}
\usepackage{multirow}
\usepackage{makecell}
\usepackage{hyperref}
\usepackage{float}
\usepackage{xkeyval}
\usepackage{soul}  %
\usepackage{wrapfig}
\begin{document}

\author{Xiaotian Su}
\orcid{0009-0004-0548-1576}
\affiliation{%
  \institution{ETH Zurich}
  \city{Zurich}
  \country{Switzerland}}
\email{xiaotian.su@inf.ethz.ch}

\author{Hongxin Fu}
\affiliation{%
  \institution{Beijing Normal University}
  \city{Zhuhai}
  \country{China}
}
\email{aprilf@mail.bnu.edu.cn}

\author{Xiaoyu Zhang}
\affiliation{%
  \institution{City University of Hong Kong}
  \city{Hong Kong}
  \country{China}}
\email{xiaoyu.zhang@cityu.edu.hk}

\author{April Yi Wang}
\orcid{0000-0001-8724-4662}
\affiliation{%
  \institution{ETH Zurich}
  \city{Zurich}
  \country{Switzerland}}
\email{april.wang@inf.ethz.ch}

\renewcommand{\shortauthors}{Su et al.}

\newcommand{\AW}[1]{\textcolor{blue}{\textbf{*April*}: #1}}
\newcommand{\SW}[1]{\textcolor{purple}{\textbf{*Su*}: #1}}
\newcommand{\XZ}[1]{\textcolor{orange}{\textbf{*Xiaoyu*}: #1}}
\newcommand{\tmp}[1]{\hl{#1}\xspace}

\newcommand{\tmpcite}{\tmp{[cite]}}
\newcommand{\psay}[1]{\textit{\say{#1}}}

\begin{abstract}
Computational notebooks make problem-solving visible, but typically only one notebook at a time. 
Meanwhile, in data science platforms like Kaggle, one competition can accumulate hundreds of notebooks.
Effective collection-level analysis requires characterizing recurring solution patterns across all notebooks, as well as isolating specific notebooks for closer examination and learning.
However, standard notebooks provide no common basis for this. Their workflows are nonlinear, cells declare no intent, and identical code can serve different ends, leaving hundreds of notebooks as separate documents. 
In this paper, we present \sys{}, an interactive visual analytics tool for profiling hundreds of notebooks as one collection. 
Inspired by a formative study (N = 11), \sys{} classifies every cell into one of thirteen \ac{ML} stages, turning each notebook into a stage sequence, and clusters those sequences by structure rather than by code. 
To keep the representation constant as the scope narrows from the whole collection to a single cell, \sys{} features three coordinated views---Workflow, Structural Matrix, and Detail---that appear at all four levels of granularity. 
In a within-subject study (N = 17) using two Kaggle collections of over 400 notebooks each, we observed participants answered questions about all notebooks more accurately with \sys{} (median 88\% vs.\ 50\%) while opening 80\% fewer notebooks per minute.
Notably, four of the fourteen answered it without opening a single notebook (interaction logs, N = 14). 
Participants also reported significantly lower mental demand, temporal demand, and stress with \sys{} (Holm–Bonferroni adjusted).
\end{abstract}

\newcommand{\sys}{\textsc{Notrix}}
\title{\sys{}: Understanding Machine Learning Solutions Across Computational Notebooks at Scale}


\begin{teaserfigure}
    \centering
    \includegraphics[width=\linewidth]{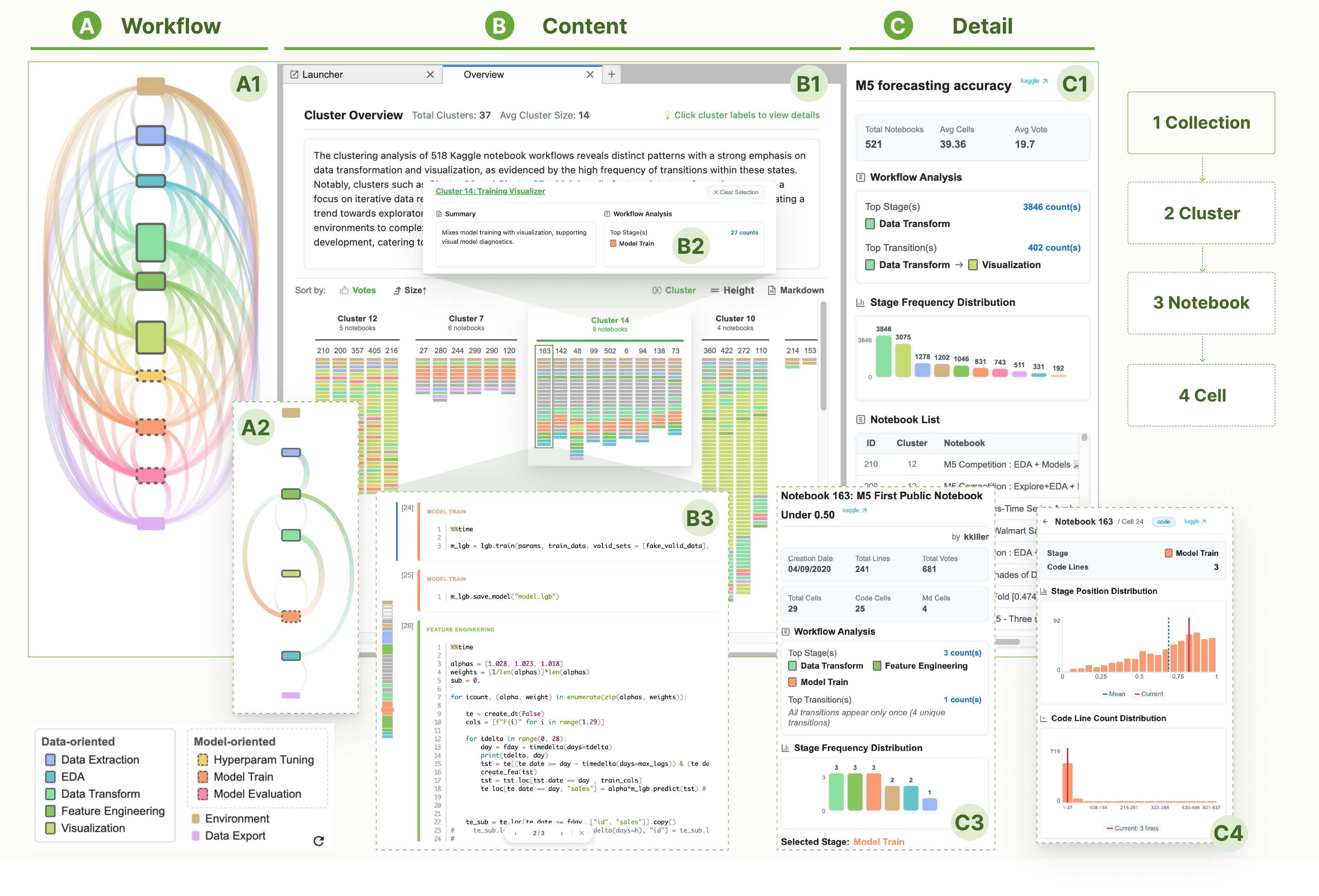}
    \caption{\sys{} comprises three coordinated regions: (A) Workflow panel, (B) Content panel with Structural Matrix, and (C) Detail Panel. These views are instantiated across four levels of a notebook collection:  (1) the whole collection, (2) a cluster, (3) a notebook, and (4) a cell, as a consistent analysis pipeline passes through. Since the same three views recur, the reading strategy a user forms over hundreds of notebooks transfers effortlessly to a single cell. This figure illustrates a representative exploration path in \sys{}: users can drill down from 521 Kaggle competition notebooks to Cluster 14, then to Notebook 163, then to one Model Train cell. Each region appears at every level where it is meaningful: the workflow view spans levels 1–3, since a single cell has no internal transitions; and cluster-level statistics are folded into the cluster overview (B2) rather than a separate detail panel.}     
    \Description{\sys{} applies one analysis consistently across four levels of a notebook collection. Three coordinated regions: (A) the workflow view, (B) the structural matrix and its content panel, and (C) the detail panel. They are instantiated at each level the analysis passes through: (1) the whole collection, (2) a cluster, (3) a notebook, and (4) a cell. Because the same three views recur, the reading strategy a user forms over hundreds of notebooks transfers unchanged to a single cell. The figure traces one such path: 521 Kaggle notebooks, one of the two collections used in our study, narrow to Cluster 14, then to Notebook 163, then to one Model Train cell. Each region appears at every level where it is meaningful: the workflow view spans levels 1–3, since a single cell has no internal transitions; and cluster-level statistics are folded into the cluster overview (B2) rather than a separate detail panel.}
    \label{fig:teaser}
\end{teaserfigure}
\maketitle

\section{Introduction}

Computational notebooks are widely used in \ac{DS} and \ac{ML} as they support iterative experimentation and flexible problem-solving workflows.
Notebook-based analysis is inherently iterative---data scientists test hypotheses, trial methods, and adjust parameters while making numerous decisions about features, models, and experimental designs \cite{chi21-forkit}.
While this process feels intuitive to the original authors, it is rarely documented in ways that external readers can easily follow.
Consequently, such variability poses challenges for understanding a single notebook~\cite{chi21-forkit, ramasamy2023visualising}, and this difficulty compounds when many notebooks address the same task.
In collection-level settings such as reviewing student submissions~\cite{tochi15-overcode, www14-codewebs}, exploring Kaggle solutions, or comparing alternative modeling strategies, users do not read each notebook line-by-line. Instead, they seek to extract higher-level insights: recurring workflows, good practices, distinctive strategies, and meaningful deviations.

Yet existing tools support only single-notebook inspection and offer little help for comparing multiple solutions or identifying patterns across them, making large-scale comparison slow, inconsistent, and cognitively demanding. Keyword-based retrieval offers limited support because it retrieves isolated cells rather than workflows and relies on users knowing the exact terminology used, which may not always be feasible in exploratory or inconsistent notebook environments.~\cite{chi18-exploration, chi21-nbsearch}.
Existing approaches for scaling code comprehension, such as VizProg~\cite{chi23-vizprog}, show promise for program snippets but do not generalize to notebooks, where reasoning is expressed through flexible workflows rather than strictly sequential code. 
The exploratory nature of notebooks introduces diverse stage ordering and structural variation, which makes it difficult to compare solutions or identify recurring patterns using tools designed for line-by-line or snippet-based comprehension.
These problems intensify at scale, as public repositories such as Kaggle host hundreds of thousands of notebooks, where manual browsing is infeasible. Addressing this challenge requires rethinking notebook tooling to support insight extraction at the scale of collections, enabling users to move beyond surface-level artifacts toward meaningful patterns in how solutions are constructed.
Such limitations can affect multiple stakeholders.
In education, instructors face growing grading demands and must evaluate not only correctness but also reasoning and coding practices, yet without collection-level overviews, important patterns are easily missed, risking inconsistent feedback~\cite{tcs24-automated, chi25-CPVis}.
Learners, such as those on Kaggle, benefit from examining peer submissions, but sifting through dozens of notebooks to identify transferable ideas slows iteration and delays learning gains.
Beyond education, recruiters must compare performance and originality across many notebooks, often without support for summarizing diverse coding styles and solution strategies, thus increasing the risk of overlooking important signals in hiring decisions.

Given these scenarios, we ask: \textbf{How can we support users in efficiently exploring, identifying, and comparing workflow patterns across hundreds of computational notebooks?} 
To answer this RQ, we introduce \sys{}, a computational notebook clustering and visualization system designed to support comprehension and comparison across large collections of Jupyter notebooks. 
Our approach is guided by prior work on workflow abstraction and provenance visualization, which emphasize the value of making analytic structure and dependencies explicit~\cite{ramasamy2023visualising, vds19-albireo}. Building on these insights, \sys{} summarizes notebook structure along thirteen predefined \ac{ML} stages~\cite{Code4ML}, such as feature engineering and model evaluation, and renders inter-cell dependencies using dynamic graph representations.
This design makes the execution order and reasoning patterns visible, enabling users to trace how outcomes are produced.
Although designed for \ac{ML} tasks, our methodology can generalize: the core process of classifying code into workflow stages and clustering based on these patterns can be adapted to other domains, such as data analysis, software engineering, or scientific simulation, by redefining the stage set to match the new context.
More broadly, providing structured, predefined signals (e.g., workflow labels) mitigates vocabulary mismatch and helps retrieve relevant notebooks and code cells more reliably, thus reducing reliance on precise query terms~\cite{chi21-nbsearch}.

To evaluate the effectiveness of our approach, we conducted a within-subject user study (N=17).
The results show that \sys{} significantly improves the participants’ ability to make sense of Jupyter notebook collections while reducing their workload. We further highlight how structure-based design principles can generalize beyond \ac{ML} notebooks to other domains of computational practice.
Our contributions are as follows:
\begin{itemize}
    \item An interactive visualization system for moving from notebook-level to collection-level comprehension, including:
    (1) workflow visualizations for quick structural overview,
    (2) a matrix view with multiple ordering schemes for workflow pattern exploration, and
    (3) synchronized multi-notebook navigation for detailed comparison.
    \item A technical pipeline that includes:
    (1) a fine-tuned model for ML workflow classification (F1 = 92.36\%), and
    (2) a structural clustering algorithm based on workflow transitions for comparing solution variations.
    \item A user study with 17 participants showing that \sys{} improves users’ ability to make sense of collection-level notebooks while easing their experience by considerably reducing their workload.
\end{itemize}

\section{Related Work}

\subsection{Large-Scale Computational Notebook Exploration}
Research has identified key challenges in computational notebooks, including fragmented structures~\cite{Grotov2022}, limited support for retrieval~\cite{li2024unlockinginsightssemanticsearch,siddik2025systematicliteraturereviewsoftware}, and difficulties in cross-workflow comparison~\cite{11016521}. 
Visualization has emerged as an effective approach to address these issues and enhance the interpretability of notebooks.
Initial efforts focused on revealing the internal structure and reasoning logic of single notebooks. 
These systems adopt diverse representations beyond the linear execution order, such as clustering semantically related cells~\cite{vds19-albireo}, constructing semantic graphs linking text, code, and outputs~\cite{chi25-interlink}, or visualizing temporal provenance to reconstruct non-linear workflows~\cite{Loops}. 
Another direction emphasizes comparison, traceability, and documentation~\cite{bhatAspirationsPracticeModel2023}.
These systems extend the notebook environment by visualizing AutoML pipelines for side-by-side evaluation~\cite{PipelineProfiler}, binding code parameters to outputs for experimental traceability~\cite{B2}, or embedding user insights directly into workflows through sketch-based annotation~\cite{InkSight}. 

Despite these advancements, most existing systems are designed for single-notebook analysis or lack alignment across standardized analytical stages~\cite{chiea24-supernova, tian2025noteflow, uist25-flowco}. 
A few recent tools have started to address this broader need. 
For instance, EDAssistant \cite{edassistant23} facilitates semantic search across collections but is confined to the \ac{EDA} phase, failing to provide traceability across the full \ac{ML} pipeline, from data cleaning to model evaluation.
NBSearch \cite{chi21-nbsearch} enhances code snippet retrieval through natural language but focuses on fragment-level exploration rather than holistic workflow comparison.
Our work bridges this gap by introducing a visualization system that aligns multiple notebooks through standardized analytical stages, enabling holistic workflow comparison and pattern discovery across collections.

\subsection{Program Variation Analysis at Scale}
Prior work on large-scale code comprehension has explored multiple ways of organizing and comparing programs, with visualization playing a central role in revealing similarities and differences. Three broad themes emerged: content-based clustering, output-based clustering, and temporal analysis.

\textbf{Content-based clustering} compares programs by their internal representation, 
ranging from structural similarity with \ac{AST} edit distances \cite{www14-codewebs, huang2013syntactic} to modern semantic embeddings like CodeBERT~\cite{feng-etal-2020-codebert} and GraphCodeBERT~\cite{guo2021graphcodebertpretrainingcoderepresentations}
In education, these representations enabled tools such as CFlow, which groups functionally related student submissions to help instructors identify common patterns and errors at scale~\cite{LAS24-cflow}.
\textbf{Output-based clustering} prioritize program behavior over form.
In engineering, this manifests as evaluating AutoML pipelines by their model performance~\cite{ieee19-automl}.
In education, dynamic analysis provides a similar perspective by executing programs on shared inputs and grouping them by outputs or traces. 
OverCode~\cite{tochi15-overcode}, for example, clusters submissions with equivalent value traces, and later work extends this to dynamic equivalence~\cite{pldi18-automated} and web programming contexts through SPARK~\cite{yang2025spark}. 
Such approaches help reveal functionally equivalent solutions, recurring strategies, and common misconceptions.
\textbf{Temporal analysis} tracks code evolution. 
In software engineering, provenance-tracking systems visualize how architectures are incrementally developed and refactored~\cite{racs22-provenance}. 
In education, systems such as VizProg~\cite{chi23-vizprog} capture fine-grained edit histories and visualize trajectories of student solutions in real time, exposing patterns in problem-solving processes and supporting timely pedagogical intervention.

Despite these advances, existing taxonomies offer limited support for comparing computational notebooks at the level of the analytical workflow. 
Notebook collections vary not just in syntax or output, but in how stages are sequenced, prioritized, or iterated.
To address this gap, we contribute a \emph{structural clustering} approach that groups notebooks by workflow patterns. 
Instead of focusing only on content or output equivalence, we capture procedural organization and stage transitions, enabling collection-level analysis of methodological diversity and common patterns in \ac{ML} workflows.

\subsection{Sequential Pattern Mining and Visualization}
Extensive research has examined how to make sequential patterns interpretable through visualization. 
\citet{survey} provide a comprehensive survey and classify event sequence visualizations into five major forms: \textit{chart-based, timeline-based, hierarchy-based, Sankey-based} and \textit{matrix-based}. 

\textit{Chart-based} representations such as bar charts, line charts, or scatter plots are frequently used to summarize event distributions or compare performance features~\cite{ViDX}. 
\textit{Timeline-based} approaches arrange events in temporal order, often enriched with icons and encodings to highlight attributes~\cite{StoryFlow}.
\textit{Hierarchy-based} visualizations (e.g., trees) have been applied to reveal branching structures in sequential data, such as motion tracking~\cite{MotionFlow}, and are also widely adopted in visualizing data science workflows in Jupyter notebooks~\cite{Rehman,ramasamy2023visualising}. 
\textit{Sankey-based} approaches represent another major category of sequential pattern visualization, which include two main types. 
Standard directed node-link graphs~\cite{CarePre} have inspired a number of variants, including interaction networks~\cite{Barnes2016DataDriven}, progress networks~\cite{10.1145/3441636.3442366,MiningStudentActivity}, and interaction graphs~\cite{10.1007/978-3-319-93843-1_24}, which have been particularly influential in the education domain for analyzing students’ programming behaviors and difficulties. 
Traditional Sankey diagrams~\cite{Frequence} also have two notable variants—chord diagrams and arc diagrams~\cite{ViSeq}, which provide alternative encodings for forward and backward transitions.
Finally, \textit{matrix-based} visualizations~\cite{MatrixWave} offer scalable representations for comparing frequent event subsequences and managing overlap across large datasets. 

These visualization techniques illustrate the diversity of approaches to representing sequential data and highlight design trade-offs between scalability, interpretability, and domain-specific adaptation. 
Our work extends this line of research by adapting sequential pattern visualizations to computational notebooks, aligning heterogeneous analytical stages across multiple workflows to enable holistic comparison and pattern discovery.

\section{Design Iteration: From Cell-Based Clustering to Structure-Based Clustering}
\label{sec:design}
Understanding multiple notebooks requires more than inspecting isolated cells. It requires making sense of workflow structure across an entire collection. This section presents how our design evolved through iterative exploration grounded in user feedback. 

\subsection{Iteration 1: Early Prototype}
\begin{figure*}
    \centering
    \includegraphics[width=\linewidth]{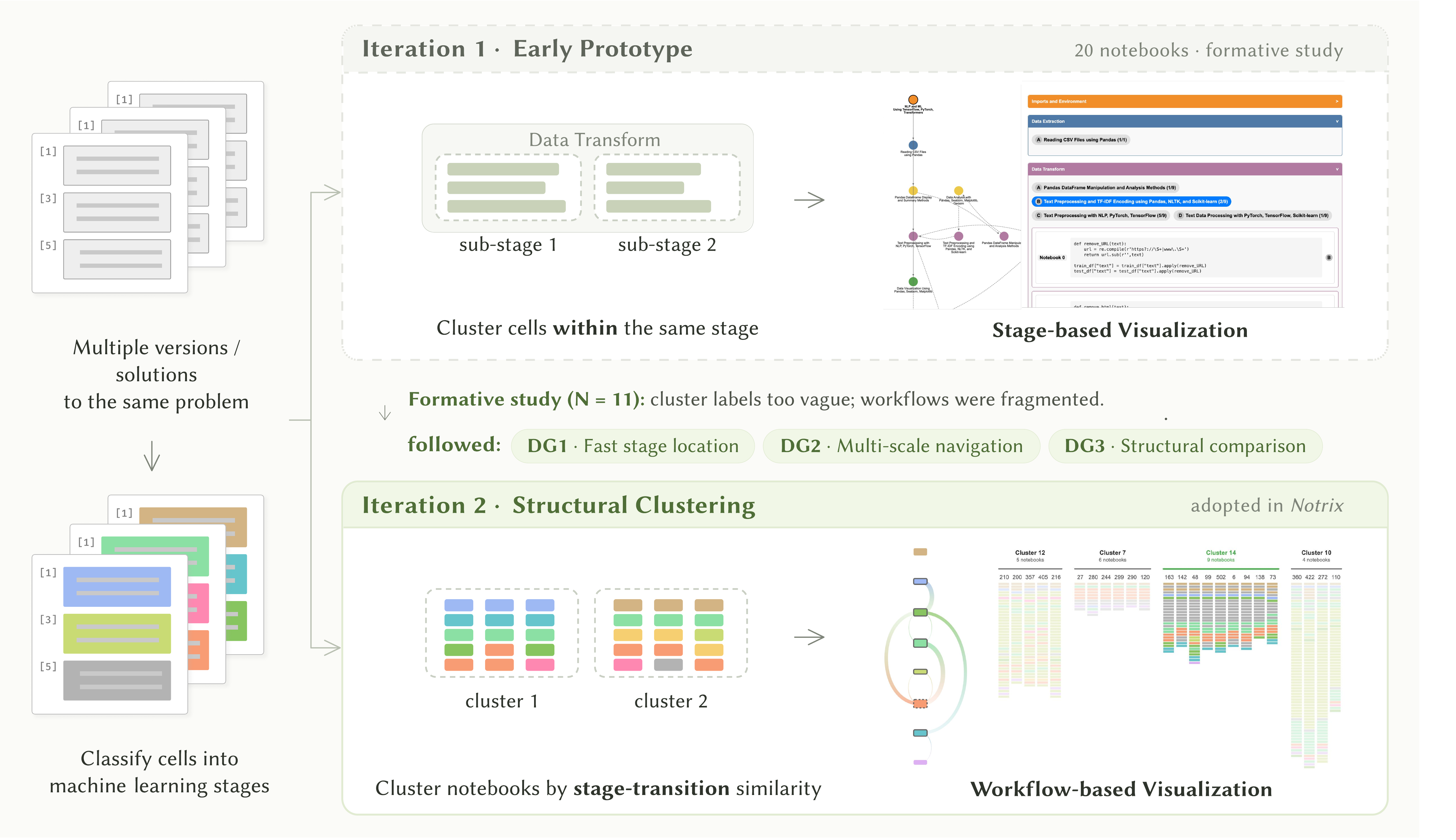}
    \caption{Design iteration from cell-based to structural clustering. Both iterations operate on the same input: multiple solutions to one problem, with every cell classified into one of \ac{ML} stages. Iteration 1 clustered cells within a stage to expose content variation. A formative study (N = 11) found the resulting fragments too vague and too disjointed to represent a notebook as a whole. Iteration 2 therefore targets structural variation, clustering whole notebooks by their stage-transition sequences so that process patterns, rather than isolated code fragments, become the unit of comparison. This is the granularity adopted in \sys{}.}
    \Description{Design iteration from cell-based to structural clustering. Both iterations operate on the same input: multiple solutions to one problem, with every cell classified into one of ML stages. Iteration 1 clustered cells within a stage to expose content variation. A formative study (N = 11) found the resulting fragments too vague and too disjointed to represent a notebook as a whole. Iteration 2 therefore targets structural variation, clustering whole notebooks by their stage-transition sequences so that process patterns, rather than isolated code fragments, become the unit of comparison. This is the granularity adopted in Notrix.}
    \label{fig:iteration}
\end{figure*}

Our first design iteration asked: \emph{How can we cluster notebooks based on cell content?} 
We applied this approach to a collection of 20 notebooks from the same Kaggle competition. 
The first prototype took the cell rather than the notebook as its unit. Every cell in the collection
was classified into an \ac{ML} stage. 
Cells carrying the same stage were then pooled across notebooks and clustered into finer groups, which we call
substages: cells labelled \emph{Data Transform}, for example, separated into text processing, dataframe manipulation, and similar groupings. The interface presented this catalogue as a stage-based panel that listed the cells gathered under each substage (Figure \ref{fig:iteration}). 
To evaluate this design, we conducted a formative study with 11 participants. 
Each participant explored two sets of 20 notebooks: once by manually browsing them with a standard viewer, and once by using our prototype. 
This comparison helped us examine how clustering changed their strategies for making sense of notebook collections.

In the baseline condition, participants often fell back on ad-hoc strategies such as \texttt{Ctrl+F} across notebooks or relying on the first well-documented notebook, leading to frustration and reduced confidence among three participants. 
The prototype improved on this in one respect. Stage and substage titles gave participants somewhere to start, and they reported knowing \psay{exactly where to look}. 
Past that first step it did not hold up.
Because cells under one substage came from many notebooks, they no longer described any of them: three participants abandoned clusters after encountering too many disparate cells, and two found the labels too vague to tell approaches apart.

As a result, the system revealed fragments of workflows but did not support reasoning about notebook-level strategies.
These findings underscored a key insight: to compare solutions, users need to see a notebook as a \emph{whole} rather than as a set of disjointed
fragments and to see the collection before choosing which notebooks to open.

\subsection{Iteration 2: Toward Structural Clustering}
In response, our second iteration shifted focus from clustering cells to clustering notebooks by their overall structure. 
Instead of asking how many ways individual code snippets can be written, we asked: \emph{How many different workflow strategies can solve the same problem?}  
Structural clustering captures differences in stage order, emphasis, omissions, and transitions. For example, some workflows spend more effort on exploration, whereas others move quickly to modeling. Some interleave evaluation throughout the process, whereas others evaluate only at the end. Some omit feature engineering entirely.
These patterns reflect strategic differences rather than algorithmic ones.
Two notebooks may use different classifiers but still follow the same overall workflow. Conversely, two notebooks may use the same model but differ substantially in how they prepare data or iterate on their designs.
By foregrounding structural variation, the design helps users compare notebooks at the level of approach rather than isolated fragments, revealing how solutions diverge, converge, or deviate across a collection while still allowing drill-down into the underlying code when finer-grained interpretation is needed.

Guided by these findings, we derived three design goals: 

\textbf{DG1: Enable fast location of workflow stages.} 
In the baseline of Iteration 1, participants often relied on ad-hoc keyword searches or scrolling through entire notebooks to locate specific stages like model evaluation or feature engineering. 
These strategies were inefficient and left participants uncertain about where relevant content was located within or across notebooks. 
The first prototype answered this directly: with cells grouped under named stages, participants reported
knowing \psay{exactly where to look}. This was the part of that design that worked, and it survived the redesign unchanged. 
Through automated per-cell stage classification and visual workflow cues, \sys{} makes stages explicit, enabling readers to quickly locate and navigate to relevant sections.

\textbf{DG2: Balance comprehensiveness with navigability.} 
Participants became overwhelmed when browsing more than a handful of notebooks, often abandoning exploration prematurely or making quick judgments without considering the full set. 
Our prototype amplified this issue: opening a cluster revealed long lists of disparate cells, which several participants scrolled through and quickly abandoned. 
\sys{} therefore aims to balance comprehensiveness with navigability by supporting multi-scale navigation, from the full collection to clusters to individual notebooks, as users drill down into the details they need.

\textbf{DG3: Enable meaningful structural comparison.} Grouping was only helpful when it reflected coherent strategies. 
In Iteration 1, participants found cluster labels too vague or repetitive to guide comparison, and noted that the fragments displayed failed to represent the workflow as a whole. 
\sys{} thus seeks to support structural comparison by improving the quality of clustering and labeling so that users can trust groups as indicative of meaningful notebook-level strategies rather than arbitrary partitions.

\section{\sys{}}
\subsection{Overview}
\sys{} supports exploration of notebook collections created for the same problem, such as the hundreds of public submissions to a Kaggle competition. It presents the collection at four levels of scope: the full collection (L1), a notebook cluster (L2), an individual notebook (L3), and a cell (L4).
A preprocessing pipeline classifies each cell by \ac{ML} stage, represents each notebook as a sequence of stages, and groups notebooks by workflow structure rather than code similarity. The interface presents this structure through three coordinated views: an arc-based workflow view, a structural matrix, and a detail panel. These views remain consistent across the four levels, while their contents update to reflect the selected scope.

\subsection{\sys{} Interface}
\begin{figure*}[htbp]
    \centering
    \begin{subcaptionbox}{Unclustered Matrix with Markdown cells displayed\label{fig:matrix-a}}[0.32\textwidth]
        {\includegraphics[width=\linewidth]{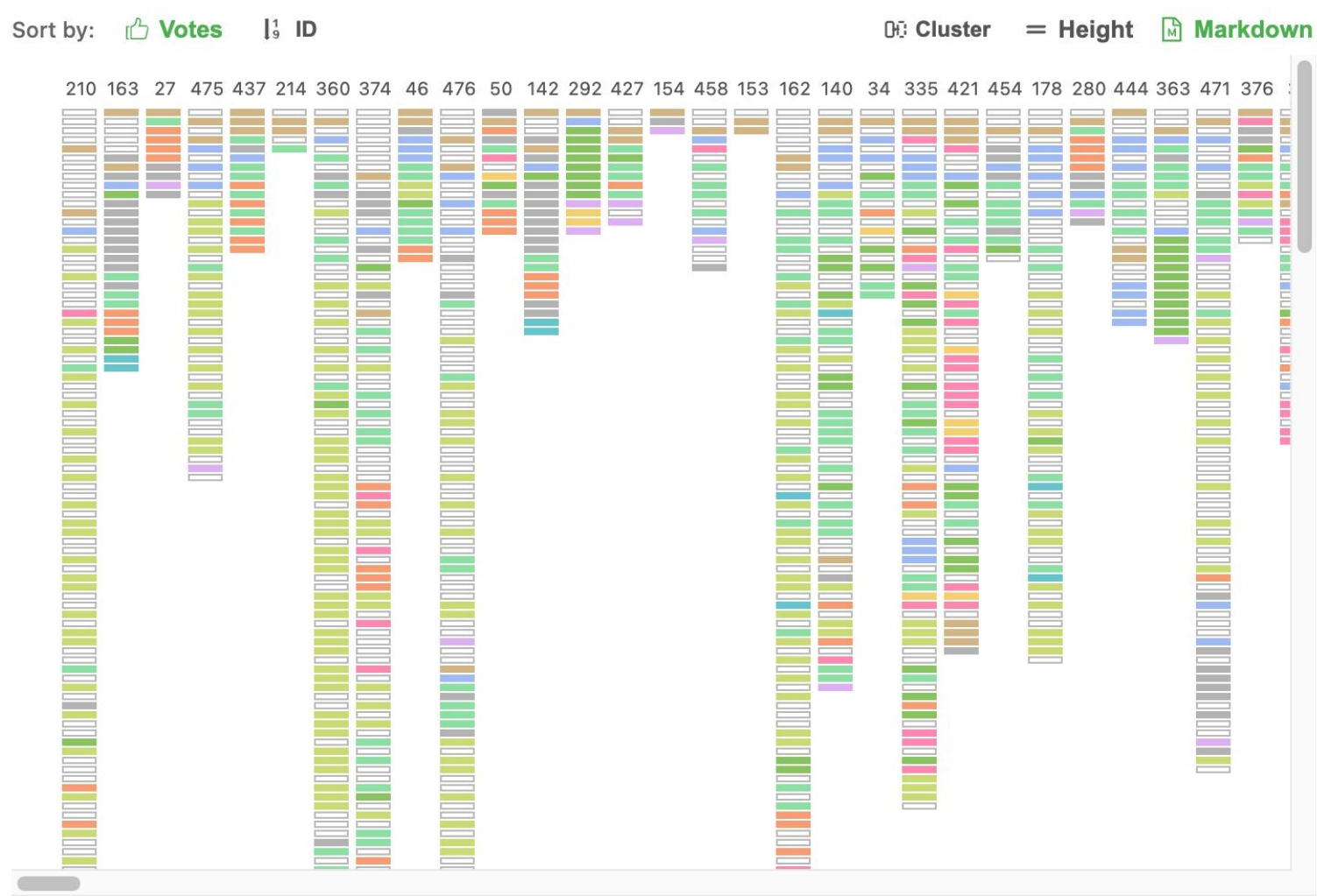}}
    \end{subcaptionbox}
    \hfill
    \begin{subcaptionbox}{Clustered Matrix with Markdown cells hidden\label{fig:matrix-b}}[0.32\textwidth]
        {\includegraphics[width=\linewidth]{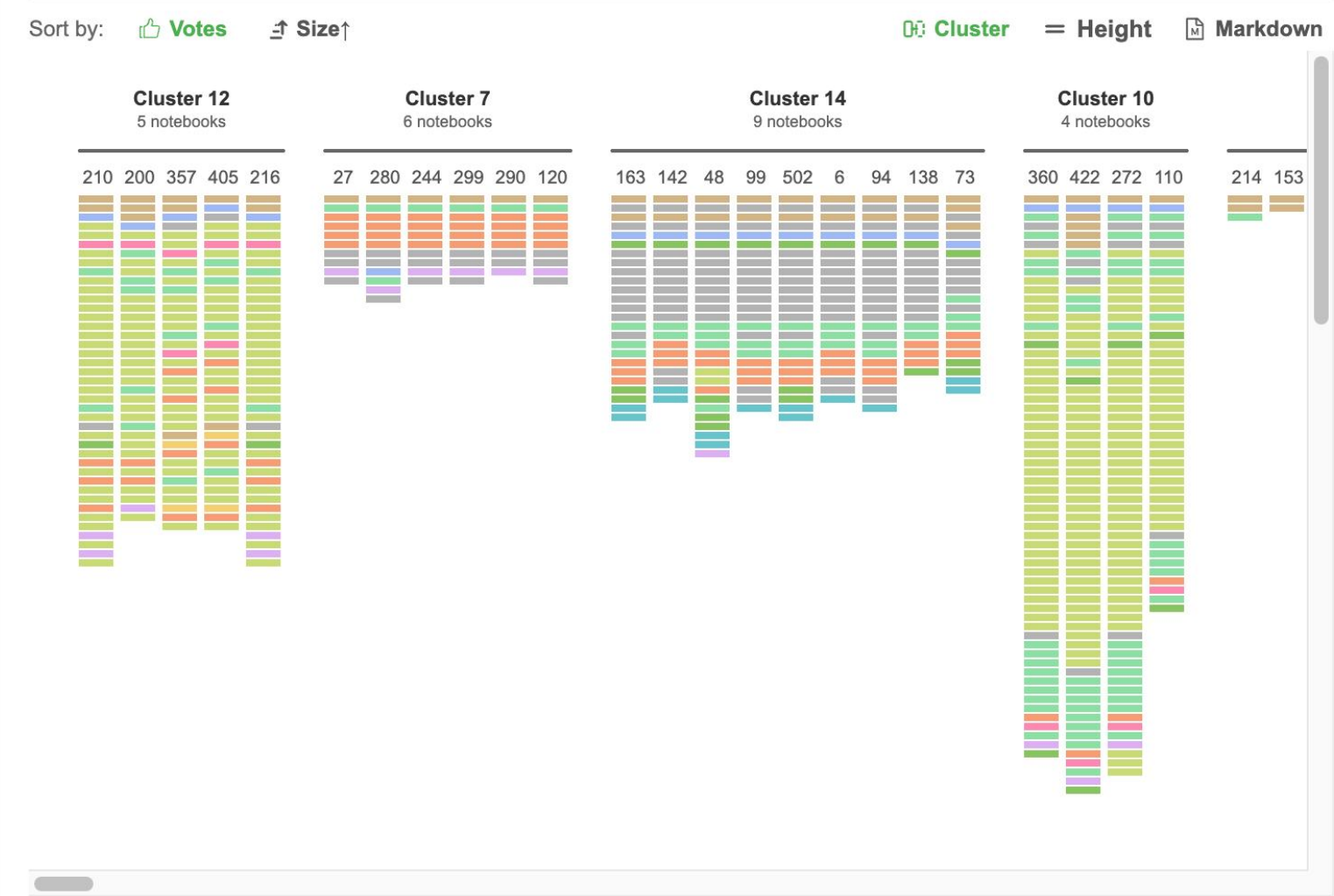}}
    \end{subcaptionbox}
    \hfill
    \begin{subcaptionbox}{Clustered Matrix with Markdown cells hidden in dynamic height mode\label{fig:matrix-c}}[0.32\textwidth]
        {\includegraphics[width=\linewidth]{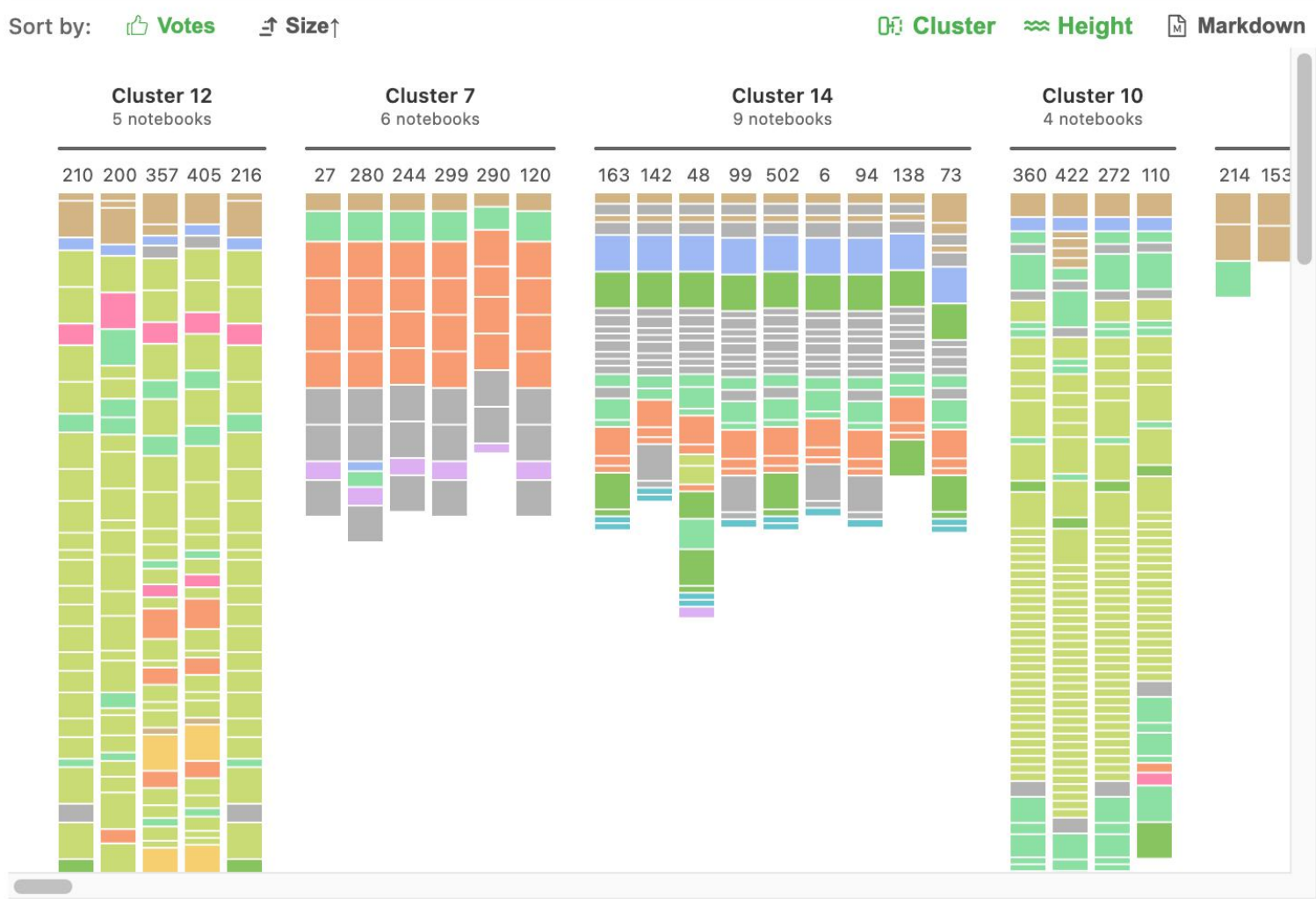}}
    \end{subcaptionbox}
    
    \caption{Structural Matrix with three display modes to reveal workflow patterns.}
    \Description{Three screenshots labelled (A), (B), and (C) show the Matrix View under different configurations, progressing from less organized to more structurally clear. Image (A) shows an unclustered matrix with both Markdown and code cells displayed, where the overall structure appears dense and harder to compare. Image (B) shows a clustered matrix with Markdown cells hidden, reducing clutter and making the underlying code structure easier to distinguish. Image (C) builds on this by applying dynamic height mode, where cell heights reflect code line counts, further clarifying structural differences across notebooks.}
    \label{fig:matrix}
\end{figure*}

Based on the design goals, we built \sys{}, a JupyterLab extension that supports interactive exploration and structural comparison of computational notebooks. 
Figure~\ref{fig:teaser} provides an overview of the system with multiple coordinated views. 
The Workflow View (Figure~\ref{fig:teaser}.A) summarizes stage frequencies and transitions, enabling users to quickly locate and navigate to stages of interest while providing a compact overview that scales across many notebooks. 
The Structural Matrix (Figure~\ref{fig:teaser}.B1--B3) groups notebooks into clusters and aligns their stage sequences. Users can reorder notebooks by different criteria and inspect cluster summaries to compare strategies while preserving collection-level overview.
The Notebook View (Figure \ref{fig:teaser}.B4) supports cell-level inspection and split-screen comparison to verify structural hypotheses and inspect differences. 
Finally, the Detail View (Figure \ref{fig:teaser}.C) provides context-aware summaries and metadata that consolidate evidence for structural comparison.

\subsubsection{Arc-Based Workflow Visualization}
The Workflow View helps users identify where ML stages typically occur, how they transition, and how these patterns differ across notebook groups.
It represents workflows as an arc diagram, a visualization well suited to sequential data because it reveals ordered relationships and recurring subsequences~\cite{1173155}.
To help users scan large collections without tracing every notebook independently (DG2), users can first examine aggregate workflow patterns across the full collection (Figure~\ref{fig:teaser}.A1).
Aggregation reveals dominant workflow patterns but can obscure individual variations; users can therefore drill down to clusters (Figure~\ref{fig:teaser}.A2) and notebooks (Figure~\ref{fig:teaser}.A3) using the same visual encoding.

The view comprises stage blocks, whose heights encode stage frequency, and transition edges, whose thicknesses encode transition frequency. Stage colors, drawn from the Pastel scheme of CARTOColors\footnote{https://carto.com/carto-colors/}, remain consistent across levels to help users track stages during drill-down.
Transition edges are rendered as cubic Bézier curves, with greater curvature for larger vertical separations, and use source-to-target color gradients to indicate direction. These encodings support comparisons of workflow structure across levels (DG3).

Hovering over or selecting a stage highlights its incoming and outgoing transitions, helping users trace its role in the workflow. Users can also reorder stages through drag-and-drop to reduce clutter or or arrange the view according to an analysis-specific mental model.

\subsubsection{Structural Matrix}
The Structural Matrix (Figure~\ref{fig:matrix}) provides a compact overview for navigating and comparing notebook structures (DG2, DG3).
Each column represents a notebook, with cells follow execution order. 
Code cells appear as filled rectangles and Markdown cells as hollow rectangles with gray outlines, enabling users to compare notebook length and code--documentation patterns across the collection (DG1).

Users can order notebooks by index or Kaggle vote count and clusters by size. Hovering over a cell reveals its metadata, while clicking opens the corresponding notebook and cell in the Notebook View.
Notebooks can be clustered by stage sequences (Figure~\ref{fig:matrix}.B).
Selecting a cluster highlights it, updates the workflow view (Figure~\ref{fig:teaser}.A2), and shows the cluster details (Figure~\ref{fig:teaser}.B2-1) above the Matrix.
Users can also hide Markdown cells to focus on executable structure (Figure~\ref{fig:matrix}.B) or scale cell height by lines of code (Figure~\ref{fig:matrix}.C).

\subsubsection{Multi-Scale Notebook Navigation and Comparison}
The Notebook View (Figure~\ref{fig:teaser}.B4) enables users to move from comparing notebook structures to inspecting individual cells.
At the \textbf{collection} level, users can open multiple notebooks side by side (Figure~\ref{fig:splitscreen}) and synchronize their scrolling to compare corresponding workflow regions. 
Panels can also be locked or navigated independently, supporting both aligned and exploratory comparisons (DG3).
At the \textbf{notebook} level, a compact minimap summarizes its structure and indicates the current viewport, helping users remain oriented in long notebooks (DG2) and navigate directly to relevant stages (DG1).
At the \textbf{cell} level, a matching vertical color bar and text label identify each cell's stage, allowing users to connect collection-level patterns with their underlying code and documentation.

\subsubsection{Context-Adaptive Information Panel}
The Detail View (Figure~\ref{fig:teaser}.C) supports meaningful structural comparison by adapting its summaries and metadata to the current selection (DG3). 
The side panel provides three levels of detail.
\textbf{Collection Detail} (Figure~\ref{fig:teaser}.C1) summarizes the full collection statistics and a stage frequency bar chart. 
The notebook list mirrors the Matrix ordering; hovering over a notebook in the Matrix highlights the corresponding list entry.
\textbf{Notebook Detail} (Figure~\ref{fig:teaser}.C3) summarizes a selected notebook statistics with metadata such as its Kaggle link, author, creation date, and line counts.
\textbf{Cell Detail} (Figure~\ref{fig:teaser}.C4) presents fine-grained metadata for selected cells with two diagnostic histograms for the stage position distribution and the code line count distribution.

\subsection{\sys{} Pipeline}
\begin{figure*}
    \centering
    \includegraphics[width=\linewidth]{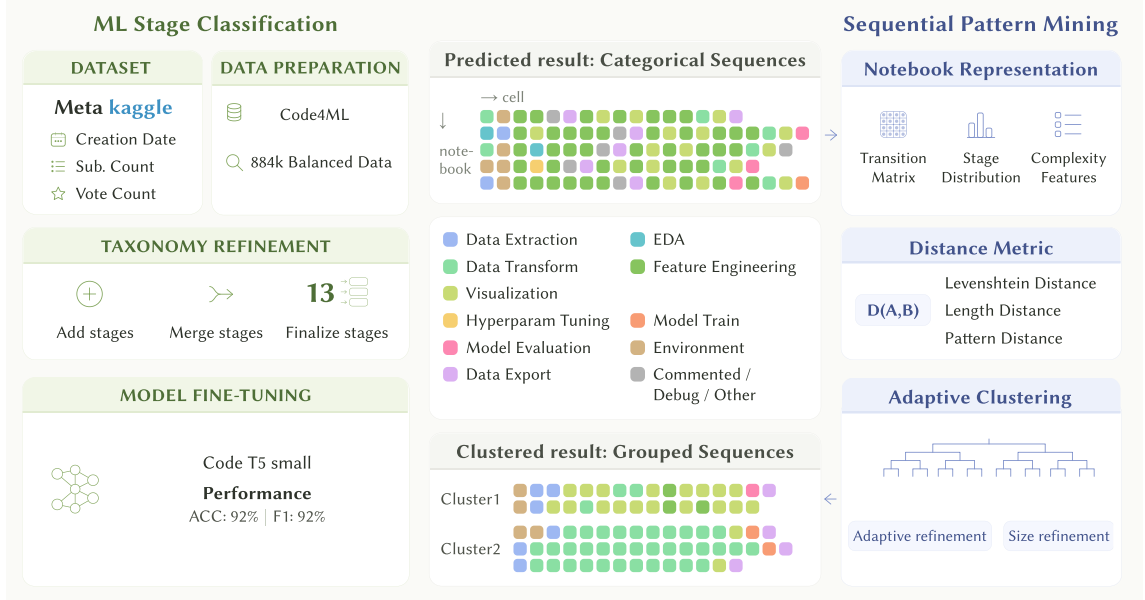}
    \caption{End-to-end pipeline for ML stage classification and sequential pattern mining of computational notebooks. The framework proceeds in two phases. \textbf{(a) ML stage classification (left).} Candidate notebooks are selected from Meta Kaggle and cell-level annotation is obtained from Code4ML, from which a class-balanced corpus of 884k cells is constructed. The source taxonomy is refined through three operations and produces a final set of 13 stages. A CodeT5-small encoder is fine-tuned on this corpus for cell-level stage prediction, reaching 92\% accuracy and 92\% F1 on the held-out split. \textbf{(b) Sequence construction (middle).} Applying the classifier to every cell converts each notebook into a categorical sequence, in which rows correspond to notebooks and columns to cell order; colors denote predicted stages (legend, middle center). Commented, Debug, and Other are analysed as distinct stages but share a single grey encoding, as none of them denotes a substantive workflow step. \textbf{ (c) Sequential pattern mining (right).} Each sequence is represented by its stage-transition matrix, marginal stage distribution, and complexity features. Pairwise dissimilarity combines three distance metrics and an adaptive hierarchical clustering procedure, groups notebooks into coherent workflow archetypes (bottom middle), revealing recurring authoring patterns across the corpus.}
    \Description{End-to-end pipeline for ML stage classification and sequential pattern mining of computational notebooks. The framework proceeds in two phases. \textbf{(a) ML stage classification (left).} Candidate notebooks are selected from Meta Kaggle and cell-level annotation is obtained from Code4ML, from which a class-balanced corpus of 884k cells is constructed. The source taxonomy is refined through three operations and produces a final set of 13 stages. A CodeT5-small encoder is fine-tuned on this corpus for cell-level stage prediction, reaching 92\% accuracy and 92\% F1 on the held-out split. \textbf{(b) Sequence construction (middle).} Applying the classifier to every cell converts each notebook into a categorical sequence, in which rows correspond to notebooks and columns to cell order; colors denote predicted stages (legend, middle center). Commented, Debug, and Other are analysed as distinct stages but share a single grey encoding, as none of them denotes a substantive workflow step. \textbf{ (c) Sequential pattern mining (right).} Each sequence is represented by its stage-transition matrix, marginal stage distribution, and complexity features. Pairwise dissimilarity combines three distance metrics and an adaptive hierarchical clustering procedure, groups notebooks into coherent workflow archetypes (bottom middle), revealing recurring authoring patterns across the corpus.}
    \label{fig:pipeline}
\end{figure*}

The \sys{} algorithmic pipeline transforms raw notebook collections into structured, interactive visualizations to fulfill our design goals (Figure~\ref{fig:pipeline}). 
To enable fast location of workflow stages (DG1), we first apply automated ML stage classification to each code cell, providing immediate semantic labels. 
To balance comprehensiveness with navigability (DG2), we cluster notebooks by workflow patterns and establish multi-level navigation hierarchies from collection to cluster to individual notebook to cell level. 
Finally, to enable meaningful structural comparison (DG3), we extract multi-faceted similarity features that capture both sequential patterns and workflow complexity.

\subsubsection{Unit of analysis}
\sys{} uses the notebook cell as its unit of analysis because cells are author-defined structural and execution boundaries: they divide the work, execute independently, and retain their associated outputs. Labelling at this level avoids imposing arbitrary code segments and preserves traceability, since readers can inspect each assigned stage directly against the corresponding code and output. Higher-level notebook and cluster patterns are then formed by aggregating these labelled cells, allowing every claim to be traced back to the underlying code.

\subsubsection{Dataset and Stage Taxonomy}
To construct a fine-tuning dataset for the code classification model, we evaluated publicly available resources, including DASWOW~\cite{ese23-workflow} and Code4ML~\cite{Code4ML}. We selected Code4ML because of its scale and diversity. It contains human-annotated Python snippets labeled with 13 high-level ML stages and approximately 80 second-level categories.

We refined its taxonomy to better reflect practical ML workflows, drawing on feedback from the first design iteration and literature on ML development and collaboration (see Table~\ref{tab:labels} in the Appendix). First, we promoted \textit{Feature Engineering} from a secondary category to a primary stage because of its established role in the ML lifecycle. 
We also introduced \textit{Commented} code as a first-level label, using rule-based detection of single- and multi-line comments to help the model learn their distinct patterns.
We further revised the \textit{Model Interpretation} and \textit{Debug} categories. Because \textit{Model Interpretation} was rare and was not consistently treated as a separate workflow stage in prior studies, we removed it and reassigned affected cells to the nearest preceding non-\textit{Model Interpretation} cell, or to the nearest subsequent cell when no preceding cell existed. 
Because \textit{Debug} cells, typically characterized by print statements, can occur throughout a workflow, we first identified them as \textit{Debug} and then assigned each the primary label of the preceding cell while retaining \textit{Debug} as a sub-label.

\subsubsection{Supervised Fine-Tuning for Stage Classification}
We removed duplicate code blocks before splitting the dataset to prevent data leakage and improve efficiency.
To meet \texttt{codet5-small}'s 512-token limit\footnote{https://huggingface.co/Salesforce/codet5-small}, we removed cells longer than approximately 2,000 characters, leaving 1.8M samples (Appendix~\ref{appendix:classes}). 
We replaced any occurrences of T5's end-of-sequence token, \texttt{</s>}, in code cells with \texttt{<slash\_s>} to prevent them from being interpreted as sequence boundaries during training.
Balanced sampling produced 1.1M examples across 13 classes (85,000 per class), split 80/10/10 into training, validation, and test sets. 
The fine-tuned model achieved 92.39\% accuracy and a 92.36\% F1-score (Appendix~\ref{appendix:clf}).

\subsubsection{Structural Representation of Notebook Sequences}
To analyze coding behavior at a higher level, we introduce an \ac{HMM}-inspired pipeline that clusters notebook stage sequences using probabilistic representations, a multi-metric distance, and adaptive hierarchical refinement. 
The method yields homogeneous clusters while controlling cluster granularity and size. 
Each notebook is modeled as a temporal sequence of categorical labels capturing the user’s trajectory. 
Following the symbolic time series framework~\cite{fournier2017survey}, we cluster by transition dynamics to reveal common workflows and recurring patterns.

To enhance interpretability, we exclude the categories \textit{Commented}, \textit{Debug}, and \textit{Other}, including sublabels like \textit{define variables}, which can occur arbitrarily and do not reflect coherent workflow stages. 
The clustering and visualization operate on ten core categories, yielding more meaningful and discriminative behavioral groupings.

\paragraph{Single Notebook Representation}
Let a notebook be a discrete sequence $A=(a_1,\dots,a_T)$ over $N$ stages, ($a_t\in\{0,\dots,N-1\}$), the algorithm summarizes $A$ with three complementary representations: (1) an empirical transition matrix $T_A\in\mathbb{R}^{N\times N}$ (in our case $N=10$), obtained by counting adjacent state transitions and converting counts to probabilities with Laplace smoothing to address sparsity;
(2) a state distribution $\pi_A\in\mathbb{R}^N$, given by the relative frequency of each stage;
(3) a 15-dimensional complexity feature vector $\mathbf{x}_A$ such as sequence length and entropy (see Appendix~\ref{appendix:hmm} for all features). 

\paragraph{Pairwise Distance for Notebook Sequences}
Using the above features, we defined a multi-metric distance measure ($D$) comprising six complementary, normalized metrics ($\mathcal{M}$) to quantify similarity between any two sequences (e.g., $A$ and $B$): (1) transition matrix distance ($d_{\text{transition}}$), (2) state distribution distance ($d_{\text{state}}$), (3) complexity feature distance ($d_{\text{complexity}}$), (4) normalized Levenshtein edit distance ($d_{\text{edit}}$),  (5) pattern distance ($d_{\text{pattern}}$), and (6) length distance ($d_{\text{length}}$). 
The overall distance is a weighted sum with configurable, non-negative weights that sum to one:
\begin{equation}
D^{(s)}(A,B)
= \sum_{m \in \mathcal{M}} w^{(s)}_{m}\, d_{m}(A,B),
\qquad
\sum_{m \in \mathcal{M}} w^{(s)}_{m}=1,\;\; w^{(s)}_{m}\!\ge\!0, 
\end{equation}

where $\mathcal{M}=\{\text{transition},\text{state},\text{complexity}, \text{edit},\text{pattern}, \text{length}\}$, $d_m$ denotes the normalized component distance for metric $m$, $w^{(s)}_m$ is the weight used in scenario $s$.

\subsubsection{Multi-metric Structural Clustering}
Given the pairwise distance matrix \(D\in\mathbb{R}^{M\times M}\) over \(M\) notebooks using the composite distance above, we first apply a multi-metric structural clustering with Ward’s method~\cite{Ward01031963} to construct a dendrogram. 
We cut the dendrogram to obtain initial clusters.
Then, we perform an adaptive refinement step to improve cluster quality. 
For each cluster, we assess cohesion via the average intra-cluster distance; if this exceeds a configurable refinement threshold, indicating internal heterogeneity, the cluster is flagged for splitting. 
We also enforce size constraints: clusters whose cardinality exceeds a specified maximum are marked for refinement to avoid overly broad groupings.
All hyperparameters are specified in a JSON configuration, and the best-tuned configuration was selected by optimizing the silhouette score~\cite{ROUSSEEUW198753} and Davies-Bouldin Index~\cite{4766909} over tuning configurations.

\section{Technical Evaluation}
We conducted a technical evaluation to assess whether the workflow clusters identified by our clustering algorithm align with human judgments of workflow similarity.
We ask: \textit{Does \sys{} produce groupings that correspond to human judgments of workflow similarity to a degree comparable to the correspondence between two independent human judgments?}

\subsection{Evaluation Sets and Annotation Procedure}
We constructed six evaluation sets across three competitions, with two independently sampled sets per competition. 
Each set contained 18 notebooks drawn from four \sys{} clusters within the same competition, yielding 108 notebooks in total.
For each evaluation set, the 18 notebooks were presented as notebook-stripe visualizations on a Miro board.
Their positions were randomly shuffled, and the \sys{} cluster assignments were hidden. Annotators saw the visualization together with its legend explaining the visual encodings.

Two external researchers with backgrounds in computer science, neither of whom is an author of this paper, independently annotated all six sets.
Both annotators received the same grouping instruction: partition the 18 notebook stripes into exactly four mutually exclusive groups by placing notebooks with similar workflow patterns together. 
They were asked to base their judgments on similarities and differences visible in the workflow visualization, including the sequence and composition of workflow elements represented by the legend. 
Every notebook had to be assigned to exactly one group. 
The annotators performed the task independently and could not inspect or discuss one another's assignments.

\subsection{Comparison Metrics}
For each evaluation set, we measured: (1) human--human agreement by comparing the partitions produced by the two annotators, and (2) human--algorithm agreement by comparing the \sys{} partition separately with each annotator's partition.

We report three complementary agreement measures for the above two comparisons. 
Best-match accuracy constructs a contingency table between two partitions and finds the maximum-overlap one-to-one matching between their clusters, leaving any surplus clusters unmatched. This provides an intuitive measure of direct cluster correspondence. 
Adjusted Rand Index (ARI) measures agreement in whether pairs of notebooks are assigned to the same or different clusters, correcting for agreement expected by chance. 
Normalized Mutual Information (NMI) measures the information shared by the two partitions, normalized by the arithmetic mean of their entropies. 
For human--human agreement, we compared the cluster assignments of the two annotators within each evaluation set, producing one value for each metric per set. 
For human--algorithm agreement, we compared \sys{} separately with each annotator, producing two values for each metric per set.

\subsection{Results}
Across the six evaluation sets, human--algorithm correspondence was descriptively close to human--human correspondence across all three measures. 
Best-match accuracy was 63.9\% for human--algorithm comparisons and 60.2\% for human--human comparisons. 
ARI was 0.2865 versus 0.2693, respectively, while NMI was 0.5239 versus 0.5271. Thus, human--algorithm agreement was 3.7\% higher in best-match accuracy and 0.0172 higher in ARI, while human--human agreement was 0.0032 higher in NMI. No metric therefore showed a substantial or consistent advantage for the human--human comparison.

We interpret these results as evidence that \sys{} captures workflow distinctions that are recognizable to independent human observers. This does \emph{not} imply that \sys{} recovers a uniquely correct taxonomy, nor that human disagreement by itself establishes that the grouping task is intrinsically ambiguous. 
Rather, under the same visual representation and four-group constraint, the partition produced by \sys{} exhibits approximately the same level of correspondence with an independent human partition as two independently produced human partitions exhibit with one another.
The moderate absolute agreement also highlights an important property of this evaluation task: notebooks may admit more than one reasonable organization depending on which aspects of their workflows an observer emphasizes. Accordingly, we use human--human agreement as a reference point for the level of correspondence achievable under independent grouping, rather than treating either annotator as ground truth.

\section{User Study}
To assess how \sys{} supports comprehension of notebook collections, we conducted a within-subject study (N = 17). 
Each participant completed two task sets, one with \sys{} and one with the baseline; condition order and collection assignment were counterbalanced. 
\sys{} aims to preserve close reading of a single notebook and extend comprehension beyond it, to comparisons of several notebooks, and to a collection as a whole. 
Therefore, the evaluation examined both what the system preserves and what it extends across three scopes: a single notebook, five notebooks sampled from the collection, and all notebooks.
All questions had verifiable answers obtainable directly from the notebooks (Appendix~\ref{appendix:comprehension}). 
We measured accuracy, logged interactions, and assessed usability and NASA-TLX questionnaires after each condition.

\subsection{Method}
\subsubsection{Participants}
We recruited 17 participants (P1–P17; 8 male, 8 female, and 1 who preferred not to disclose; aged 20–28, M = 24.4, SD = 2.9). 
The group included graduate students, five doctoral students with teaching assistant experience (P4, P6, P14, P15, P17), one participant preparing course materials for the upcoming semester (P12), and several experienced practitioners. 
All participants had prior exposure to machine learning techniques.

\subsubsection{Materials}
The study used three Kaggle competitions and two interfaces.
We selected Kaggle because it provides large public collections of independently authored notebooks addressing the same prediction problem, enabling controlled comparison of alternative ML workflows.
\paragraph{Competitions.} We restricted selection to competitions held between 2020 and 2025, excluding earlier and classic tasks (e.g., Titanic, House Prices) whose solutions many participants would already have encountered, and prioritized competitions with high participation so that a large body of public notebooks was available. 
Three competitions met these criteria: 
(C1) M5 Forecasting Accuracy\footnote{\url{https://www.kaggle.com/competitions/m5-forecasting-accuracy}} (2020), 
(C2) Home Credit--Credit Risk Model Stability\footnote{\url{https://www.kaggle.com/competitions/home-credit-credit-risk-model-stability}} (2024), and (C3) American Express--Default Prediction\footnote{\url{https://www.kaggle.com/competitions/amex-default-prediction}} (2022). 
C1 and C2 were used for the comprehension tasks and C3 for the warm-up. 
Each participant saw one competition per condition, with competition counterbalanced against condition. Table~\ref{tab:competitions} reports more details.
All figures in this paper are drawn from C1.

\paragraph{Interfaces.}
Both conditions ran inside the same JupyterLab environment, so they differed only in the interface under study and not in the surrounding tooling.
The baseline supported everything
the tasks require: a list of all notebooks in the collection (Figure~\ref{fig:baseline}.A), each notebook's full content with a table of
contents, and per-notebook metadata (Figure~\ref{fig:baseline}.B).

\subsubsection{Procedure}
We conducted a 90-minute, within-subjects online user study.
Participants first reported their prior expertise in \ac{DS}, computational notebooks, and Kaggle.
We then explained the study procedure. 
We provided a structured system demonstration followed by a warm-up task.
Participants then completed a 20-minute comprehension task for both systems. 
After completing tasks for each system, participants submitted a post-task survey. 
Each session concluded with a semi-structured interview to elicit feedback on the system’s utility for workflow discovery.

\subsubsection{Measures}
Objective measures were comprehension accuracy and task completion time. We logged all interactions in the \sys{} condition and notebook openings in the baseline; because both interfaces open notebooks in separate tabs, this event was directly comparable.
After each condition, participants completed a 7-point survey combining NASA-TLX workload items with SUS-adapted usability items (Appendix~\ref{appendix:survey}). 
After using \sys{}, they also rated feature usefulness and overall perception. 
Interviews examined feature value and trade-offs, while participants with teaching or grading experience discussed how grouping and similarity views might support notebook review.

\subsubsection{Analysis}
We compared the two conditions on comprehension accuracy, questionnaire items, and interaction-log measures, and analysed the post-session interviews to explain the patterns these measures revealed.

\paragraph{Survey and task performance.} 
Because the design was within-subjects, we tested paired differences for normality using Shapiro–Wilk tests, applying paired t-tests when normality held and Wilcoxon signed-rank tests otherwise.
Effect sizes were Cohen's $d_z$ for t-tests and $r=Z/\sqrt{N}$, respectively.
We controlled family-wise error using Holm–Bonferroni correction: questionnaire items were treated as one family, while the three task scopes were corrected separately for accuracy and time. 
We report Holm-adjusted p-values and medians with interquartile ranges.

\paragraph{Interaction logs.} 
Logs were available for 14 of 17 participants; three failed due to local browser issues.
We assigned each event to one of four levels from the interface state: 
\emph{collection} when the matrix was shown in a non-clustered ordering; 
\emph{cluster} when it was ordered by structural similarity or a cluster had been selected; 
\emph{notebook} when a notebook tab was in focus; 
and \emph{cell} when a cell detail was open. 
A cross-level transition is a consecutive pair of events assigned to different levels. 
Between-condition analyses use notebook openings, a shared measure by both interfaces; all other interaction measures apply only to \sys{}.

\paragraph{Interview.} 
Following a consensus-coding protocol~\cite{mcdonald2019reliability}, two researchers independently coded five of the 17 transcripts and merged their codes into a preliminary codebook, resolving discrepancies by discussion. 
They then grouped related codes into themes and applied the finalized codebook to the remaining transcripts.

\subsection{Results}
\begin{figure*}[t]
    \centering
    \includegraphics[width=\linewidth]{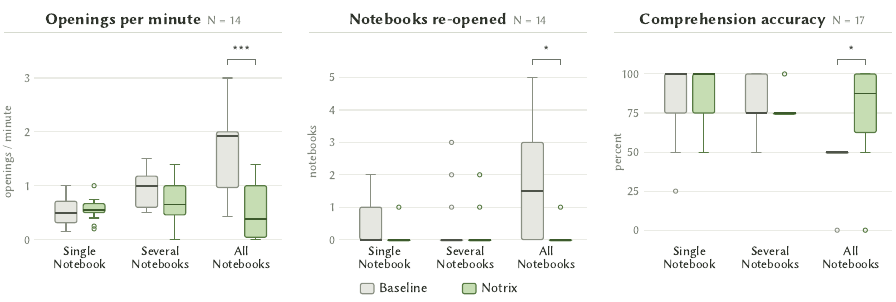}
    \caption{Notebook opening and comprehension accuracy by task scope. The two opening panels cover the 14 participants whose sessions were logged; comprehension accuracy covers all 17. On All Notebooks, participants using \sys{} opened notebooks at 0.39 per minute against 1.92 with the baseline ($d_z = -1.52$, $p_{\mathrm{Holm}} = .0002$), re-opened a median of 0 notebooks against 1.5 ($d_z = -0.88$, $p_{\mathrm{Holm}} = .017$), and answered more accurately (median 87.5\% against 50.0\%; $r = .74$, $p_{\mathrm{Holm}} = .013$). Significance is based on paired tests with Holm–Bonferroni correction for all items (*$p<0.05$, **$p<0.01$, ***$p<0.001$).}
    \Description{Notebook opening and comprehension accuracy by task scope. The two opening panels cover the 14 participants whose sessions were logged; comprehension accuracy covers all 17. On All Notebooks, participants using \sys{} opened notebooks at 0.39 per minute against 1.92 with the baseline ($d_z = -1.52$, $p_{\mathrm{Holm}} = .0002$), re-opened a median of 0 notebooks against 1.5 ($d_z = -0.88$, $p_{\mathrm{Holm}} = .017$), and answered more accurately (median 87.5\% against 50.0\%; $r = .74$, $p_{\mathrm{Holm}} = .013$). Significance is based on paired tests with Holm–Bonferroni correction for all items (*$p<0.05$, **$p<0.01$, ***$p<0.001$).}
    \label{fig:task}
\end{figure*}

\subsubsection{Reading and Comprehension Differed at the Collection Scope}
On All Notebooks tasks participants using \sys{} opened a third as many notebooks as with the baseline and answered the question more accurately.
They opened notebooks at 0.39 per minute against 1.92 ($p_{\mathrm{Holm}} = .0002$), a median of 3 notebooks against 10 ($p_{\mathrm{Holm}} = .001$), and re-opened a median of 0 against 1.5 ($p_{\mathrm{Holm}} = .017$). 
Accuracy on the same task rose from a median of 50.0\% to 87.5\% ($p_{\mathrm{Holm}} = .013$).
Four of the 14 answered without opening a single notebook; with the baseline the same four had opened between 6 and 17 notebooks and each scored 50\%, and with \sys{} they scored 83\%, 100\%, 88\%, and 50\%. 
Higher accuracy is usually bought with more effort. It was not here.
With \sys{}, Mental demand (mean 3.24 against 4.29, $p_{\mathrm{Holm}} =.019$), temporal demand (3.06 against 4.35, $p_{\mathrm{Holm}} = .007$) and stress (2.24 against 3.76, $p_{\mathrm{Holm}} = .017$) were all lower with \sys{}. 
Effort required was lower on average but not significantly so (3.65 against 4.53, $p_{\mathrm{Holm}} = .10$).
Participants' own ratings point the same way: they judged \sys{} more effective for exploring alternatives (mean 5.82 against 3.53 on a 7-point scale, $p_{\mathrm{Holm}}= .006$) and for identifying similar solutions (6.41 against 3.76, $p_{\mathrm{Holm}} = .003$), 13 of 17 rating it higher on each.

On the Single Notebook task, participants opened notebooks at the same rate with both (0.55 per minute with \sys{} against 0.50, $p_{\mathrm{Holm}} = .83$), and both performed ceiling on accuracy (median 100\%, 71\% of participants scored full marks with \sys{} and 59\% with the baseline). 
The structure \sys{} placed above the notebook therefore left close reading of one notebook unaffected.

\subsubsection{Cross-Level Notebook Navigation}
Across interaction logs, participant accounts, subjective ratings, and task performance, our findings indicate that \sys{} supported smooth transitions between overview and detail. 
Interaction logs first demonstrated that participants actively navigated across these different scales.
Interaction logs from the 14 participants whose sessions were recorded show all four levels in active use. 
Thirteen worked at every level during the tasks. The median number of cross-level transitions was 24 per session, indicating that they repeatedly moved between overview and detail rather than using \sys{} only as a static summary.
Activity was spread across the hierarchy rather than concentrated at one level: 22.8\% of interactions were at the collection level, 14.6\% at the cluster level, 50.4\% at the notebook level, and 12.2\% at the cell level. The cluster level was reached deliberately rather than incidentally, in that 11 of the 14 either re-ordered the matrix by structural similarity or selected a cluster (Figure~\ref{fig:time} in Appendix~\ref{appendix:navigation}).

Participants’ qualitative accounts help explain how the different views supported these navigation and comparison practices.
At the notebook level, participants found the workflow visualization helpful for understanding structure and locating stages of interest. 
They described the stage-based representation as matching how they naturally conceptualize \ac{ML} processes. 
For example, P3 noted that the visualization made it \psay{easier to identify specific sections}, P16 suggested integrating this view into Kaggle because it would \psay{help focus on one part of the notebook}.
For a small set of notebooks, the side-by-side view supported direct comparison of individual notebooks. 
P3 said it \psay{helps to see whether two notebooks are similar,} a benefit also highlighted by P14 and P15.
And 14 of the 17 participants used this feature (video recordings).
At the collection level, the workflow visualization and matrix view helped participants compare notebook structure at scale. 
P15 described the broad overview as \psay{really helpful for grasping the overall structure,} and others similarly valued the matrix for comparing notebooks of varying complexity (P6, P10, P16).

\begin{figure*}
    \centering
    \includegraphics[width=\linewidth]{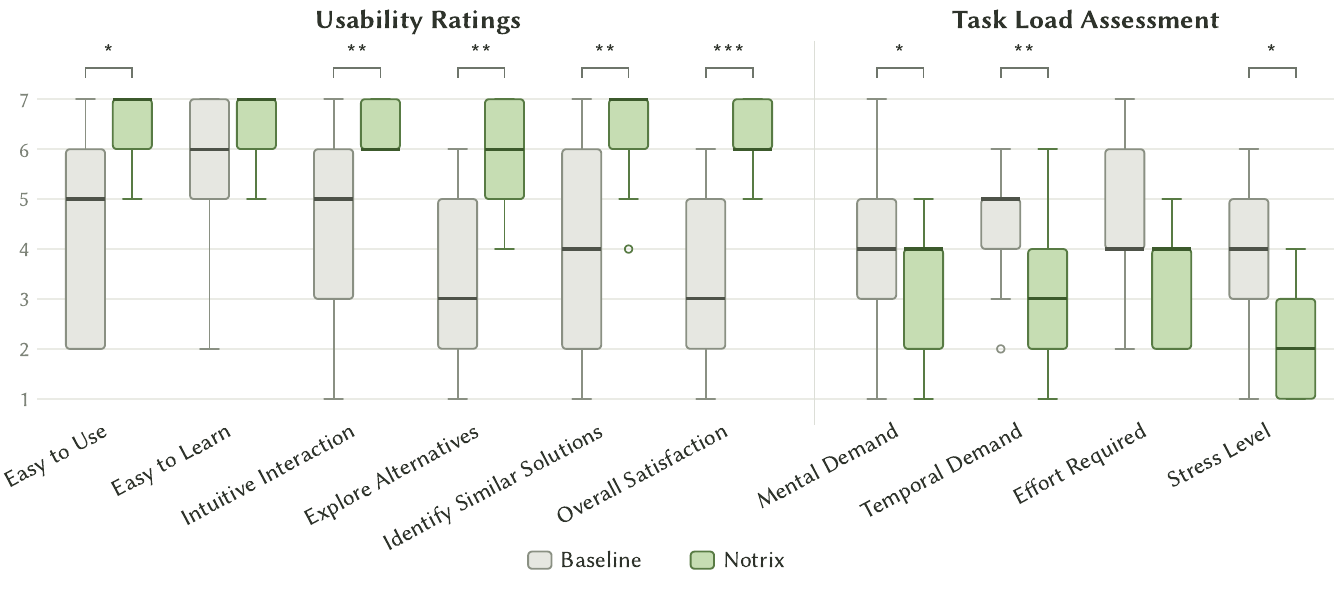}
    \caption{Subjective ratings of usability and task load by condition. \sys{} was rated significantly higher on five of six usability items and lower on three of four task-load items. Ratings were given on 7-point scales (N = 17, within-subject): usability (left; higher is better) and task load (right; lower is better). Significance is based on paired tests with Holm–Bonferroni correction for all items (*$p<0.05$, **$p<0.01$, ***$p<0.001$), reflecting within-participant change, not the separation of the two distributions.}
    \Description{Subjective ratings of usability and task load by condition. \sys{} was rated significantly higher on five of six usability items and lower on three of four task-load items. Ratings were given on 7-point scales (N = 17, within-subject): usability (left; higher is better) and task load (right; lower is better). Significance is based on paired tests with Holm–Bonferroni correction for all items (*$p<0.05$, **$p<0.01$, ***$p<0.001$), reflecting within-participant change, not the separation of the two distributions.}
    \label{fig:comparison}
\end{figure*}

These perceived benefits were also reflected in subjective ratings and task performance.
Participants rated \sys{} significantly higher than the baseline on \say{easy to use} and \say{intuitive to interact with} (Figure~\ref{fig:comparison}, left). 
Participants also gave \sys{} higher ratings on its \say{effective in exploring alternatives} and \say{identifying similar solutions} (both $p_{\mathrm{Holm}} < .01$), together with higher overall satisfaction ($p_{Holm} < .001$). 
Feature-specific ratings further reinforced these findings: Workflow Visualization received a median usefulness rating of 6, with most responses in the agree-to-strongly-agree range. 
These benefits were also reflected in task performance. 
These results suggest that \sys{} supported more systematic exploration of large notebook collections without overwhelming users.

Our evaluation demonstrates that \sys{} transforms how users approach large notebook collections. 
Participants found \sys{} easier to use and less cognitively demanding than baseline approaches, with improvements in usability metrics and reductions in mental workload. 
Through triangulation of interaction logs, quantitative metrics, and qualitative insights, we show how workflow-centric design shifts notebook analysis from tedious manual browsing to efficient, systematic exploration.

\subsubsection{Stage labels and clusters were used to decide what to open}
Participants valued stage classification and structural clustering that enabled more systematic exploration of large notebook collections without inspecting each notebook individually. 
Stage labels made the structure of a notebook legible from the outside: P2 said \psay{this automated tagging of the code cells help a lot,} and P9, who named it the most useful feature, said it gave \psay{a comprehensive overview of entire notebooks.} Clustering carried the same judgement across notebooks, surfacing relationships that browsing one notebook at a time does not reveal. P13 described using it to \psay{trace how other people improved on that original notebook} by looking at which cluster the top-voted notebooks fell into, and P16 found it \psay{really helpful to identify the workflow and identify the similarity of notebooks,} an observation shared by P1, P3, P8, and P11.

\begin{figure}
  \centering
  \includegraphics[width=\linewidth]{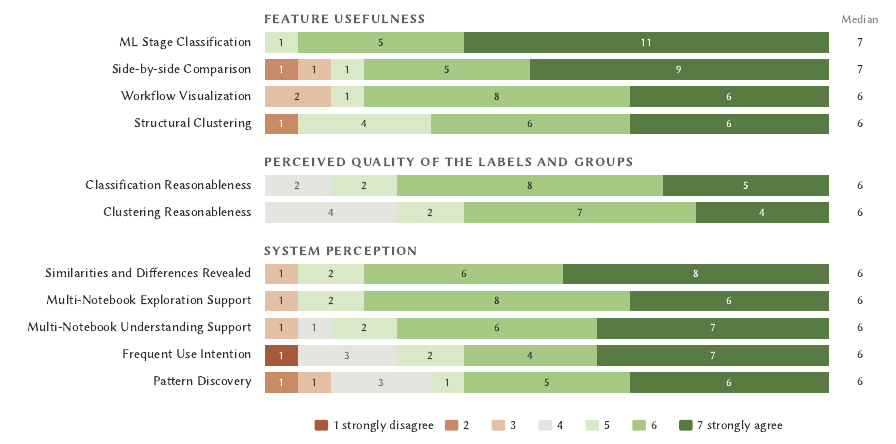}
    \caption{Post-task questionnaire evaluation for \sys{}, on a 7-point Likert scale (1 = strongly disagree, 7 = strongly agree; N = 17). Numbers inside each segment give the count of participants at that point, and rows are ordered by mean within each block. All four features reached a median of 6 or 7 and were rated 5 or above by at least 15 of the 17 participants; ML Stage Classification received no rating below 5 and the most top-scale responses (11 of 17). See Appendix~\ref{appendix:notrix} for full question wording.}
    \Description{Post-task questionnaire evaluation for \sys{}, on a 7-point Likert scale (1 = strongly disagree, 7 = strongly agree; N = 17). Numbers inside each segment give the count of participants at that point, and rows are ordered by mean within each block. All four features reached a median of 6 or 7 and were rated 5 or above by at least 15 of the 17 participants; ML Stage Classification received no rating below 5 and the most top-scale responses (11 of 17). See Appendix~\ref{appendix:notrix} for full question wording.}
  \label{fig:ratings}
\end{figure}

The ratings follow the same order (Figure~\ref{fig:ratings}). Stage classification was the highest-rated feature of the system (median 7, 11 of 17 selecting strongly agree) and structural clustering follows closely (median 6). Perceived reasonableness of both outputs reached a median of 6. 
The interaction logs show that this acceptance changed behavior. 
On the All Notebooks task, four of the fourteen logged participants (P7, P8, P13, P17) answered without opening a single notebook reached a median accuracy of 85\%. With the baseline, those same four had opened between 6 and 17 notebooks each and all scored 50\%.

\subsubsection{Uses beyond the study tasks}
Working with a class requires instructors to reason across a collection of notebooks before acting on any individual one. This collection-level view supports not only the assessment of submitted work, but also the preparation of instructional materials.
Participants extended the system’s analytical structures to these teaching activities. Five of them had graded coursework as teaching assistants (P4, P6, P14, P15, P17), and we asked them a further set of questions about assessment. 
Their responses highlighted two complementary operations: partitioning a set of submissions into meaningful groups and aligning an individual submission with a reference.

P17 described both operations. Clusters became a way to organize the grading workflow rather than merely an analytical result: \psay{I can grade students' submissions based on the cluster,} and synchronized comparison became a diff against a template answer: \psay{on the right I open my reference answer, and on the left is the student's notebook. As I scroll, it automatically aligns with the student's corresponding module, so I can directly check and grade without having to search.} 

Collection-level analysis also informed the preparation of instructional materials.
4 proposed using stage-transition statistics to guide the design of fill-in-the-blank exercises: \psay{As TAs, we usually first write a full notebook and then remove key code for students to fill in. I want to ensure the loop sequence matches most students' tendencies, so these statistics are meaningful for template design.}
Similarly, P12 used Markdown density to identify notebooks suitable for instructional explanation, noting that the feature helped reveal \psay{which notebook containing more markdown information can be used to create course materials.}

\section{Discussion}
\subsection{Challenges and Opportunities for Scalable Multi-Notebook Comprehension}
Our findings show that scaling comprehension from individual notebooks to collections introduces both cognitive and systems-level challenges. Because notebooks are fragmented, nonlinear, and highly variable in organization, naming, and documentation \cite{chi21-forkit, ramasamy2023visualising}, multi-notebook understanding cannot be supported by lexical search alone. 
Instead, it requires workflow structure to be represented as a first-class abstraction for comparison.
Participants frequently struggled to locate semantically meaningful stages, such as feature engineering or model evaluation, across notebooks. 
\sys{} addresses this challenge through a cell-level classification pipeline that maps cells to thirteen ML workflow stages and exposes those stages directly in the interface, enabling navigation by analytical function rather than surface text. Our study further highlights that scale requires multi-resolution views: users need both collection-level summaries and direct paths to notebook- and cell-level detail. This motivated our matrix-style visualization, which supports fluid movement between overview and inspection and makes large notebook collections more interpretable.

At the same time, we found that the effectiveness of such interfaces depends critically on the robustness of the intermediate semantic representations they expose. Although the workflow classifier performed well under common naming conventions, we observed systematic failures when variables and identifiers were written in German. In these cases, cells were more likely to be assigned to incorrect workflow stages, which in turn distorted the structural summaries presented in the interface. 
Addressing this limitation suggests several technical directions for future work, including incorporating more language-agnostic program representations such as execution traces, or dataflow relations; adopting multilingual or cross-lingual code representations; and exposing classifier uncertainty in the interface so that users can identify potentially unreliable annotations before relying on them in analysis.

\subsection{Multi-Notebook Understanding with Multi-Modal Representation}
While structural workflows provide a powerful lens for comparing notebooks at scale, our study also revealed important limitations of relying solely on code-derived signals.
Computational notebooks contain a variety of signals beyond source code, such as markdown commentary, intermediate outputs, and execution order that convey intent, rationale, and process. These modalities often capture aspects of analytical reasoning that are not visible in code structure alone. 

Future work could incorporate these additional modalities to capture aspects of analytical reasoning that are not visible in code structure alone.
Natural language content in markdown cells or comments conveys methodological intent and instructional scaffolding \cite{liu-etal-2021-haconvgnn-hierarchical}, suggesting that NLP-based analysis could complement code-derived stages by clarifying higher-level reasoning.
Output pattern analysis offers another avenue for validating workflow completeness and identifying expected intermediate results \cite{tochi15-overcode, yang2025spark}. 
If notebooks include execution order into metadata, they could be incorporated into workflow extraction, supporting clustering and visualization based on actual execution paths. Additionally, cells unrelated to core logic (e.g., installation, one-off experiments, reinitialization) could be down-weighted through behavioral or temporal signals. 
Cross-modal alignment linking code, text, outputs, and execution traces would enable richer, more accurate representations of workflow structure across diverse notebook styles.

\subsection{Limitations}
The study has several limitations. 
A cell spanning multiple stages receives a single label. Participants therefore requested multi-label support for long cells, a natural extension of the classifier.
The evaluation was limited in scope: all 17 participants were students, and each used both interfaces in a single session. Although several had teaching-assistant experience and commented on potential instructional uses of \sys{}, we treat these remarks as observations rather than an instructor evaluation.
Future work could evaluate \sys{} with instructors.

\section{Conclusion}
We introduced Notrix, a workflow-aware visualization system that addresses the challenge of understanding and comparing collections of Jupyter notebooks at scale. By classifying notebook cells into thirteen predefined ML stages and clustering notebooks based on workflow patterns, Notrix provides an overview-first approach to multi-notebook comprehension. Our user study with 17 participants demonstrated that Notrix significantly reduced cognitive workload and improved task efficiency and usability compared to traditional notebook browsing interfaces. Participants found the ML stage classification particularly valuable, with workflow visualization and side-by-side comparison features receiving strong positive ratings. 
The system successfully bridges the gap between single-notebook tools and collection-level analysis needs, enabling educators, learners, and practitioners to efficiently identify patterns, compare approaches, and extract insights from large notebook collections. These findings suggest that structured, workflow-aware visualizations represent a promising direction for scaling notebook comprehension in data science workflows.

\bibliographystyle{ACM-Reference-Format}
\bibliography{ref}

@ARTICLE{4766909,
  author={Davies, David L. and Bouldin, Donald W.},
  journal={IEEE Transactions on Pattern Analysis and Machine Intelligence}, 
  title={A Cluster Separation Measure}, 
  year={1979},
  volume={PAMI-1},
  number={2},
  pages={224-227},
  doi={10.1109/TPAMI.1979.4766909}}

@article{ROUSSEEUW198753,
title = {Silhouettes: A graphical aid to the interpretation and validation of cluster analysis},
journal = {Journal of Computational and Applied Mathematics},
volume = {20},
pages = {53-65},
year = {1987},
issn = {0377-0427},
doi = {https://doi.org/10.1016/0377-0427(87)90125-7},
url = {https://www.sciencedirect.com/science/article/pii/0377042787901257},
author = {Peter J. Rousseeuw}
}

@article{Ward01031963,
author = {Joe H. Ward Jr.},
title = {Hierarchical Grouping to Optimize an Objective Function},
journal = {Journal of the American Statistical Association},
volume = {58},
number = {301},
pages = {236--244},
year = {1963},
publisher = {Taylor \& Francis},
doi = {10.1080/01621459.1963.10500845},
URL = { 
    
    
        https://www.tandfonline.com/doi/abs/10.1080/01621459.1963.10500845
    

},
eprint = { 
    
    
        https://www.tandfonline.com/doi/pdf/10.1080/01621459.1963.10500845
    

}

}

@inproceedings{huang2013syntactic,
  title={Syntactic and functional variability of a million code submissions in a machine learning mooc},
  author={Huang, Jonathan and Piech, Chris and Nguyen, Andy and Guibas, Leonidas},
  booktitle={AIED 2013 Workshops Proceedings Volume},
  volume={25},
  year={2013}
}

@inproceedings{www14-codewebs,
author = {Nguyen, Andy and Piech, Christopher and Huang, Jonathan and Guibas, Leonidas},
title = {Codewebs: scalable homework search for massive open online programming courses},
year = {2014},
isbn = {9781450327442},
publisher = {Association for Computing Machinery},
address = {New York, NY, USA},
url = {https://doi.org/10.1145/2566486.2568023},
doi = {10.1145/2566486.2568023},
booktitle = {Proceedings of the 23rd International Conference on World Wide Web},
pages = {491–502},
numpages = {12},
location = {Seoul, Korea},
series = {WWW '14}
}

@article{tochi15-overcode,
author = {Glassman, Elena L. and Scott, Jeremy and Singh, Rishabh and Guo, Philip J. and Miller, Robert C.},
title = {OverCode: Visualizing Variation in Student Solutions to Programming Problems at Scale},
year = {2015},
issue_date = {April 2015},
publisher = {Association for Computing Machinery},
address = {New York, NY, USA},
volume = {22},
number = {2},
issn = {1073-0516},
url = {https://doi.org/10.1145/2699751},
doi = {10.1145/2699751},
journal = {ACM Trans. Comput.-Hum. Interact.},
month = mar,
articleno = {7},
numpages = {35}
}

@article{fournier2017survey,
  title={A survey of sequential pattern mining},
  author={Fournier-Viger, Philippe and Lin, Jerry Chun-Wei and Kiran, Rage Uday and Koh, Yun Sing and Thomas, Rincy},
  journal={Data Science and Pattern Recognition},
  volume={1},
  number={1},
  pages={54--77},
  year={2017}
}

@inproceedings{chi18-exploration,
author = {Rule, Adam and Tabard, Aur\'{e}lien and Hollan, James D.},
title = {Exploration and Explanation in Computational Notebooks},
year = {2018},
isbn = {9781450356206},
publisher = {Association for Computing Machinery},
address = {New York, NY, USA},
url = {https://doi.org/10.1145/3173574.3173606},
doi = {10.1145/3173574.3173606},
booktitle = {Proceedings of the 2018 CHI Conference on Human Factors in Computing Systems},
pages = {1–12},
numpages = {12},
location = {Montreal QC, Canada},
series = {CHI '18}
}

@INPROCEEDINGS{vds19-albireo,
  author={Wenskovitch, John and Zhao, Jian and Carter, Scott and Cooper, Matthew and North, Chris},
  booktitle={2019 IEEE Visualization in Data Science (VDS)}, 
  title={Albireo: An Interactive Tool for Visually Summarizing Computational Notebook Structure}, 
  year={2019},
  pages={1-10},
  doi={10.1109/VDS48975.2019.8973385}
}

@INPROCEEDINGS{ieee19-automl,
  author={Truong, Anh and Walters, Austin and Goodsitt, Jeremy and Hines, Keegan and Bruss, C. Bayan and Farivar, Reza},
  booktitle={2019 IEEE 31st International Conference on Tools with Artificial Intelligence (ICTAI)}, 
  title={Towards Automated Machine Learning: Evaluation and Comparison of AutoML Approaches and Tools}, 
  year={2019},
  volume={},
  number={},
  pages={1471-1479},
  doi={10.1109/ICTAI.2019.00209}
}

@article{cscw19-hai-ds,
author = {Wang, Dakuo and Weisz, Justin D. and Muller, Michael and Ram, Parikshit and Geyer, Werner and Dugan, Casey and Tausczik, Yla and Samulowitz, Horst and Gray, Alexander},
title = {Human-AI Collaboration in Data Science: Exploring Data Scientists' Perceptions of Automated AI},
year = {2019},
issue_date = {November 2019},
publisher = {Association for Computing Machinery},
address = {New York, NY, USA},
volume = {3},
number = {CSCW},
url = {https://doi.org/10.1145/3359313},
doi = {10.1145/3359313},
journal = {Proc. ACM Hum.-Comput. Interact.},
month = nov,
articleno = {211},
numpages = {24}
}

@article{aggarwal2019can,
  title={How can AI automate End-to-End data science?},
  author={Aggarwal, Charu and Bouneffouf, Djallel and Samulowitz, Horst and Buesser, Beat and Hoang, Thanh and Khurana, Udayan and Liu, Sijia and Pedapati, Tejaswini and Ram, Parikshit and Rawat, Ambrish and others},
  journal={arXiv preprint arXiv:1910.14436},
  year={2019}
}

@inproceedings{feng-etal-2020-codebert,
    title = "{C}ode{BERT}: A Pre-Trained Model for Programming and Natural Languages",
    author = "Feng, Zhangyin  and
      Guo, Daya  and
      Tang, Duyu  and
      Duan, Nan  and
      Feng, Xiaocheng  and
      Gong, Ming  and
      Shou, Linjun  and
      Qin, Bing  and
      Liu, Ting  and
      Jiang, Daxin  and
      Zhou, Ming",
    editor = "Cohn, Trevor  and
      He, Yulan  and
      Liu, Yang",
    booktitle = "Findings of the Association for Computational Linguistics: EMNLP 2020",
    month = nov,
    year = "2020",
    address = "Online",
    publisher = "Association for Computational Linguistics",
    url = "https://aclanthology.org/2020.findings-emnlp.139/",
    doi = "10.18653/v1/2020.findings-emnlp.139",
    pages = "1536--1547",
}

@misc{guo2021graphcodebertpretrainingcoderepresentations,
      title={GraphCodeBERT: Pre-training Code Representations with Data Flow}, 
      author={Daya Guo and Shuo Ren and Shuai Lu and Zhangyin Feng and Duyu Tang and Shujie Liu and Long Zhou and Nan Duan and Alexey Svyatkovskiy and Shengyu Fu and Michele Tufano and Shao Kun Deng and Colin Clement and Dawn Drain and Neel Sundaresan and Jian Yin and Daxin Jiang and Ming Zhou},
      year={2021},
      eprint={2009.08366},
      archivePrefix={arXiv},
      primaryClass={cs.SE},
      url={https://arxiv.org/abs/2009.08366}, 
}

@inproceedings{liu-etal-2021-haconvgnn-hierarchical,
    title = "{HAC}onv{GNN}: Hierarchical Attention Based Convolutional Graph Neural Network for Code Documentation Generation in {J}upyter Notebooks",
    author = "Liu, Xuye  and
      Wang, Dakuo  and
      Wang, April  and
      Hou, Yufang  and
      Wu, Lingfei",
    editor = "Moens, Marie-Francine  and
      Huang, Xuanjing  and
      Specia, Lucia  and
      Yih, Scott Wen-tau",
    booktitle = "Findings of the Association for Computational Linguistics: EMNLP 2021",
    month = nov,
    year = "2021",
    address = "Punta Cana, Dominican Republic",
    publisher = "Association for Computational Linguistics",
    url = "https://aclanthology.org/2021.findings-emnlp.381/",
    doi = "10.18653/v1/2021.findings-emnlp.381",
    pages = "4473--4485",
}

@article{csur21-ml-lifecycle,
author = {Ashmore, Rob and Calinescu, Radu and Paterson, Colin},
title = {Assuring the Machine Learning Lifecycle: Desiderata, Methods, and Challenges},
year = {2021},
issue_date = {June 2022},
publisher = {Association for Computing Machinery},
address = {New York, NY, USA},
volume = {54},
number = {5},
issn = {0360-0300},
url = {https://doi.org/10.1145/3453444},
doi = {10.1145/3453444},
journal = {ACM Comput. Surv.},
month = may,
articleno = {111},
numpages = {39}
}

@inproceedings{chi21-nbsearch,
author = {Li, Xingjun and Wang, Yuanxin and Wang, Hong and Wang, Yang and Zhao, Jian},
title = {NBSearch: Semantic Search and Visual Exploration of Computational Notebooks},
year = {2021},
isbn = {9781450380966},
publisher = {Association for Computing Machinery},
address = {New York, NY, USA},
url = {https://doi.org/10.1145/3411764.3445048},
doi = {10.1145/3411764.3445048},
booktitle = {Proceedings of the 2021 CHI Conference on Human Factors in Computing Systems},
articleno = {308},
numpages = {14},
location = {Yokohama, Japan},
series = {CHI '21}
}

@inproceedings{chi21-forkit,
author = {Weinman, Nathaniel and Drucker, Steven M. and Barik, Titus and DeLine, Robert},
title = {Fork It: Supporting Stateful Alternatives in Computational Notebooks},
year = {2021},
isbn = {9781450380966},
publisher = {Association for Computing Machinery},
address = {New York, NY, USA},
url = {https://doi.org/10.1145/3411764.3445527},
doi = {10.1145/3411764.3445527},
booktitle = {Proceedings of the 2021 CHI Conference on Human Factors in Computing Systems},
articleno = {307},
numpages = {12},
location = {Yokohama, Japan},
series = {CHI '21}
}

@article{zhang2022coral,
  title={Coral: Code representation learning with weakly-supervised transformers for analyzing data analysis},
  author={Zhang, Ge and Merrill, Mike A and Liu, Yang and Heer, Jeffrey and Althoff, Tim},
  journal={EPJ Data Science},
  volume={11},
  number={1},
  pages={14},
  year={2022},
  publisher={Springer Berlin Heidelberg}
}

@inproceedings{racs22-provenance,
author = {Burgess, Kaitlynn and Hart, Dante and Elsayed, Amr and Cerny, Tomas and Bures, Miroslav and Tisnovsky, Pavel},
title = {Visualizing architectural evolution via provenance tracking: a systematic review},
year = {2022},
isbn = {9781450393980},
publisher = {Association for Computing Machinery},
address = {New York, NY, USA},
url = {https://doi.org/10.1145/3538641.3561493},
doi = {10.1145/3538641.3561493},
booktitle = {Proceedings of the Conference on Research in Adaptive and Convergent Systems},
pages = {83–91},
numpages = {9},
location = {Virtual Event, Japan},
series = {RACS '22}
}

@inproceedings{bhatAspirationsPracticeModel2023,
  title = {Aspirations and {{Practice}} of {{Model Documentation}}: {{Moving}} the {{Needle}} with {{Nudging}} and {{Traceability}}},
  shorttitle = {Aspirations and {{Practice}} of {{Model Documentation}}},
  booktitle = {{{CHI}}},
  author = {Bhat, Avinash and Coursey, Austin and Hu, Grace and Li, Sixian and Nahar, Nadia and Zhou, Shurui and K{\"a}stner, Christian and Guo, Jin L. C.},
  year = {2023},
  doi = {10.1145/3544548.3581518},
  url = {http://arxiv.org/abs/2204.06425},
  urldate = {2023-04-20},
  archiveprefix = {arxiv},
  journal = {arXiv 2204.06425}
}

@article{Code4ML,
  title={Code4ML: a large-scale dataset of annotated Machine Learning code},
  author={Drozdova, Anastasia and Trofimova, Ekaterina and Guseva, Polina and Scherbakova, Anna and Ustyuzhanin, Andrey},
  journal={PeerJ Computer Science},
  volume={9},
  pages={e1230},
  year={2023},
  publisher={PeerJ Inc.}
}

@article{edassistant23,
author = {Li, Xingjun and Zhang, Yizhi and Leung, Justin and Sun, Chengnian and Zhao, Jian},
title = {EDAssistant: Supporting Exploratory Data Analysis in Computational Notebooks with In Situ Code Search and Recommendation},
year = {2023},
issue_date = {March 2023},
publisher = {Association for Computing Machinery},
address = {New York, NY, USA},
volume = {13},
number = {1},
issn = {2160-6455},
url = {https://doi.org/10.1145/3545995},
doi = {10.1145/3545995},
journal = {ACM Trans. Interact. Intell. Syst.},
month = mar,
articleno = {1},
numpages = {27},
}

@article{sigmod23-ml-liftcycle,
author = {Schlegel, Marius and Sattler, Kai-Uwe},
title = {Management of Machine Learning Lifecycle Artifacts: A Survey},
year = {2023},
issue_date = {December 2022},
publisher = {Association for Computing Machinery},
address = {New York, NY, USA},
volume = {51},
number = {4},
issn = {0163-5808},
url = {https://doi.org/10.1145/3582302.3582306},
doi = {10.1145/3582302.3582306},
journal = {SIGMOD Rec.},
month = jan,
pages = {18–35},
numpages = {18}
}

@inproceedings{chi23-vizprog,
author = {Zhang, Ashley Ge and Chen, Yan and Oney, Steve},
title = {VizProg: Identifying Misunderstandings By Visualizing Students’ Coding Progress},
year = {2023},
isbn = {9781450394215},
publisher = {Association for Computing Machinery},
address = {New York, NY, USA},
url = {https://doi.org/10.1145/3544548.3581516},
doi = {10.1145/3544548.3581516},
booktitle = {Proceedings of the 2023 CHI Conference on Human Factors in Computing Systems},
articleno = {596},
numpages = {16},
location = {Hamburg, Germany},
series = {CHI '23}
}

@article{ese23-workflow,
author = {Ramasamy, Dhivyabharathi and Sarasua, Cristina and Bacchelli, Alberto and Bernstein, Abraham},
title = {Workflow analysis of data science code in public GitHub repositories},
year = {2023},
issue_date = {Jan 2023},
publisher = {Kluwer Academic Publishers},
address = {USA},
volume = {28},
number = {1},
issn = {1382-3256},
url = {https://doi.org/10.1007/s10664-022-10229-z},
doi = {10.1007/s10664-022-10229-z},
journal = {Empirical Softw. Engg.},
month = jan,
numpages = {47},
}

@article{ramasamy2023visualising,
  title={Visualising data science workflows to support third-party notebook comprehension: an empirical study},
  author={Ramasamy, Dhivyabharathi and Sarasua, Cristina and Bacchelli, Alberto and Bernstein, Abraham},
  journal={Empirical Software Engineering},
  volume={28},
  number={3},
  pages={58},
  year={2023},
  publisher={Springer}
}

@inproceedings{dis24-ds-hai,
author = {Zhu, Qian and Wang, Dakuo and Ma, Shuai and Wang, April Yi and Chen, Zixin and Khurana, Udayan and Ma, Xiaojuan},
title = {Towards Feature Engineering with Human and AI’s Knowledge: Understanding Data Science Practitioners’ Perceptions in Human\&AI-Assisted Feature Engineering Design},
year = {2024},
isbn = {9798400705830},
publisher = {Association for Computing Machinery},
address = {New York, NY, USA},
url = {https://doi.org/10.1145/3643834.3661517},
doi = {10.1145/3643834.3661517},
booktitle = {Proceedings of the 2024 ACM Designing Interactive Systems Conference},
pages = {1789–1804},
numpages = {16},
location = {Copenhagen, Denmark},
series = {DIS '24}
}

@inproceedings{LAS24-cflow,
author = {Zhang, Ashley Ge and Tang, Xiaohang and Oney, Steve and Chen, Yan},
title = {CFlow: Supporting Semantic Flow Analysis of Students' Code in Programming Problems at Scale},
year = {2024},
isbn = {9798400706332},
publisher = {Association for Computing Machinery},
address = {New York, NY, USA},
url = {https://doi.org/10.1145/3657604.3662025},
doi = {10.1145/3657604.3662025},
booktitle = {Proceedings of the Eleventh ACM Conference on Learning @ Scale},
pages = {188–199},
numpages = {12},
location = {Atlanta, GA, USA},
series = {L@S '24}
}

@inproceedings{chiea24-supernova,
author = {Wang, Zijie J. and Munechika, David and Lee, Seongmin and Chau, Duen Horng},
title = {SuperNOVA: Design Strategies and Opportunities for Interactive Visualization in Computational Notebooks},
year = {2024},
isbn = {9798400703317},
publisher = {Association for Computing Machinery},
address = {New York, NY, USA},
url = {https://doi.org/10.1145/3613905.3650848},
doi = {10.1145/3613905.3650848},
booktitle = {Extended Abstracts of the CHI Conference on Human Factors in Computing Systems},
articleno = {304},
numpages = {17},
location = {Honolulu, HI, USA},
series = {CHI EA '24}
}

@article{tcs24-automated,
author = {Messer, Marcus and Brown, Neil C. C. and K\"{o}lling, Michael and Shi, Miaojing},
title = {Automated Grading and Feedback Tools for Programming Education: A Systematic Review},
year = {2024},
issue_date = {March 2024},
publisher = {Association for Computing Machinery},
address = {New York, NY, USA},
volume = {24},
number = {1},
url = {https://doi.org/10.1145/3636515},
doi = {10.1145/3636515},
journal = {ACM Trans. Comput. Educ.},
month = feb,
articleno = {10},
numpages = {43}
}

@inproceedings{uist25-flowco,
author = {Freund, Stephen N. and Simon, Brooke and Berger, Emery D. and Jun, Eunice},
title = {Flowco: Mixed-Initiative Authoring of Reliable End-to-End Data Analyses via Dataflow Graphs and LLMs},
year = {2025},
isbn = {9798400720376},
publisher = {Association for Computing Machinery},
address = {New York, NY, USA},
url = {https://doi.org/10.1145/3746059.3747636},
doi = {10.1145/3746059.3747636},
booktitle = {Proceedings of the 38th Annual ACM Symposium on User Interface Software and Technology},
articleno = {182},
numpages = {20},
location = {
},
series = {UIST '25}
}

@article{tian2025noteflow,
  title={NoteFlow: Recommending Charts as Sight Glasses for Tracing Data Flow in Computational Notebooks},
  author={Tian, Yuan and Deng, Dazhen and Yang, Sen and Zheng, Huawei and Shi, Bowen and Xiong, Kai and Yi, Xinjing and Wu, Yingcai},
  journal={arXiv preprint arXiv:2502.02326},
  year={2025}
}

@inproceedings{chi25-CPVis,
  author = {Zhang, Gefei and Ji, Shenming and Li, Yicao and Tang, Jingwei and Ding, Jihong and Xia, Meng and Sun, Guodao and Liang, Ronghua},
  title = {CPVis: Evidence-based Multimodal Learning Analytics for Evaluation in Collaborative Programming},
  year = {2025},
  isbn = {9798400713941},
  publisher = {Association for Computing Machinery},
  address = {New York, NY, USA},
  url = {https://doi.org/10.1145/3706598.3713353},
  doi = {10.1145/3706598.3713353},
  booktitle = {Proceedings of the 2025 CHI Conference on Human Factors in Computing Systems},
  articleno = {50},
  numpages = {26},
  series = {CHI '25}
}

@inproceedings{chi25-interLink,
author = {Lin, Yanna and Yang, Leni and Li, Haotian and Qu, Huamin and Moritz, Dominik},
title = {InterLink: Linking Text with Code and Output in Computational Notebooks},
year = {2025},
isbn = {9798400713941},
publisher = {Association for Computing Machinery},
address = {New York, NY, USA},
url = {https://doi.org/10.1145/3706598.3714104},
doi = {10.1145/3706598.3714104},
booktitle = {Proceedings of the 2025 CHI Conference on Human Factors in Computing Systems},
articleno = {51},
numpages = {15},
location = {
},
series = {CHI '25}
}

@ARTICLE{Loops,
  author={Eckelt, Klaus and Gadhave, Kiran and Lex, Alexander and Streit, Marc},
  journal={IEEE Transactions on Visualization and Computer Graphics}, 
  title={Loops: Leveraging Provenance and Visualization to Support Exploratory Data Analysis in Notebooks}, 
  year={2025},
  volume={31},
  number={1},
  pages={1213-1223},
  doi={10.1109/TVCG.2024.3456186}}

@ARTICLE{PipelineProfiler,
  author={Ono, Jorge Piazentin and Castelo, Sonia and Lopez, Roque and Bertini, Enrico and Freire, Juliana and Silva, Claudio},
  journal={IEEE Transactions on Visualization and Computer Graphics}, 
  title={PipelineProfiler: A Visual Analytics Tool for the Exploration of AutoML Pipelines}, 
  year={2021},
  volume={27},
  number={2},
  pages={390-400},
  doi={10.1109/TVCG.2020.3030361}}

@inproceedings{B2,
author = {Wu, Yifan and Hellerstein, Joseph M. and Satyanarayan, Arvind},
title = {B2: Bridging Code and Interactive Visualization in Computational Notebooks},
year = {2020},
isbn = {9781450375146},
publisher = {Association for Computing Machinery},
address = {New York, NY, USA},
url = {https://doi.org/10.1145/3379337.3415851},
doi = {10.1145/3379337.3415851},
booktitle = {Proceedings of the 33rd Annual ACM Symposium on User Interface Software and Technology},
pages = {152–165},
numpages = {14},
location = {Virtual Event, USA},
series = {UIST '20}
}

@ARTICLE{InkSight,
  author={Lin, Yanna and Li, Haotian and Yang, Leni and Wu, Aoyu and Qu, Huamin},
  journal={IEEE Transactions on Visualization and Computer Graphics}, 
  title={InkSight: Leveraging Sketch Interaction for Documenting Chart Findings in Computational Notebooks}, 
  year={2024},
  volume={30},
  number={1},
  pages={944-954},
  doi={10.1109/TVCG.2023.3327170}}

@ARTICLE{survey,
  author={Guo, Yi and Guo, Shunan and Jin, Zhuochen and Kaul, Smiti and Gotz, David and Cao, Nan},
  journal={IEEE Transactions on Visualization and Computer Graphics}, 
  title={Survey on Visual Analysis of Event Sequence Data}, 
  year={2022},
  volume={28},
  number={12},
  pages={5091-5112},
  doi={10.1109/TVCG.2021.3100413}}

@ARTICLE{ViDX,
  author={Xu, Panpan and Mei, Honghui and Ren, Liu and Chen, Wei},
  journal={IEEE Transactions on Visualization and Computer Graphics}, 
  title={ViDX: Visual Diagnostics of Assembly Line Performance in Smart Factories}, 
  year={2017},
  volume={23},
  number={1},
  pages={291-300},
  doi={10.1109/TVCG.2016.2598664}}

@ARTICLE{StoryFlow,
  author={Liu, Shixia and Wu, Yingcai and Wei, Enxun and Liu, Mengchen and Liu, Yang},
  journal={IEEE Transactions on Visualization and Computer Graphics}, 
  title={StoryFlow: Tracking the Evolution of Stories}, 
  year={2013},
  volume={19},
  number={12},
  pages={2436-2445},
  doi={10.1109/TVCG.2013.196}}

@ARTICLE{MotionFlow,
  author={Jang, Sujin and Elmqvist, Niklas and Ramani, Karthik},
  journal={IEEE Transactions on Visualization and Computer Graphics}, 
  title={MotionFlow: Visual Abstraction and Aggregation of Sequential Patterns in Human Motion Tracking Data}, 
  year={2016},
  volume={22},
  number={1},
  pages={21-30},
  doi={10.1109/TVCG.2015.2468292}}

@article{mcdonald2019reliability,
  title={Reliability and inter-rater reliability in qualitative research: Norms and guidelines for CSCW and HCI practice},
  author={McDonald, Nora and Schoenebeck, Sarita and Forte, Andrea},
  journal={Proceedings of the ACM on human-computer interaction},
  volume={3},
  number={CSCW},
  pages={1--23},
  year={2019},
  publisher={ACM New York, NY, USA}
}

@inproceedings{Rehman,
author = {Rehman, Mohammed Suhail},
title = {Towards Understanding Data Analysis Workflows using a Large Notebook Corpus},
year = {2019},
isbn = {9781450356435},
publisher = {Association for Computing Machinery},
address = {New York, NY, USA},
url = {https://doi.org/10.1145/3299869.3300107},
doi = {10.1145/3299869.3300107},
booktitle = {Proceedings of the 2019 International Conference on Management of Data},
pages = {1841–1843},
numpages = {3},
location = {Amsterdam, Netherlands},
series = {SIGMOD '19}
}

@article{CarePre,
author = {Jin, Zhuochen and Cui, Shuyuan and Guo, Shunan and Gotz, David and Sun, Jimeng and Cao, Nan},
title = {CarePre: An Intelligent Clinical Decision Assistance System},
year = {2020},
issue_date = {January 2020},
publisher = {Association for Computing Machinery},
address = {New York, NY, USA},
volume = {1},
number = {1},
url = {https://doi.org/10.1145/3344258},
doi = {10.1145/3344258},
journal = {ACM Trans. Comput. Healthcare},
month = mar,
articleno = {6},
numpages = {20}
}

@incollection{Barnes2016DataDriven,
  author    = {Barnes, Tiffany and Mostafavi, Behrooz and Eagle, Michael J.},
  title     = {Data-Driven Domain Models for Problem Solving},
  booktitle = {Design Recommendations for Intelligent Tutoring Systems: 
               Volume 4 -- Domain Modeling},
  editor    = {Sottilare, Robert A. and Graesser, Arthur C. and Hu, Xiangen 
               and Olney, Andrew and Nye, Benjamin and Sinatra, Anne M.},
  publisher = {U.S. Army Research Laboratory},
  year      = {2016},
  pages     = {137--146},
  isbn      = {978-0-9893923-9-6}
}

@inproceedings{10.1145/3441636.3442366,
author = {McBroom, Jessica and Paassen, Benjamin and Jeffries, Bryn and Koprinska, Irena and Yacef, Kalina},
title = {Progress Networks as a Tool for Analysing Student Programming Difficulties},
year = {2021},
isbn = {9781450389761},
publisher = {Association for Computing Machinery},
address = {New York, NY, USA},
url = {https://doi.org/10.1145/3441636.3442366},
doi = {10.1145/3441636.3442366},
booktitle = {Proceedings of the 23rd Australasian Computing Education Conference},
pages = {158–167},
numpages = {10},
location = {Virtual, SA, Australia},
series = {ACE '21}
}

@InProceedings{MiningStudentActivity,
author="Wang, Ziwei
and Jeffries, Bryn
and Koprinska, Irena",
editor="Cristea, Alexandra I.
and Walker, Erin
and Lu, Yu
and Santos, Olga C.
and Isotani, Seiji",
title="Mining Student Activity Sequences in a Large-Scale Online Programming Course",
booktitle="Artificial Intelligence in Education. Posters and Late Breaking Results, Workshops and Tutorials, Industry and Innovation Tracks, Practitioners, Doctoral Consortium, Blue Sky, and WideAIED",
year="2025",
publisher="Springer Nature Switzerland",
address="Cham",
pages="359--365",
isbn="978-3-031-99264-3"
}

@InProceedings{10.1007/978-3-319-93843-1_24,
author="McBroom, Jessica
and Yacef, Kalina
and Koprinska, Irena
and Curran, James R.",
editor="Penstein Ros{\'e}, Carolyn
and Mart{\'i}nez-Maldonado, Roberto
and Hoppe, H. Ulrich
and Luckin, Rose
and Mavrikis, Manolis
and Porayska-Pomsta, Kaska
and McLaren, Bruce
and du Boulay, Benedict",
title="A Data-Driven Method for Helping Teachers Improve Feedback in Computer Programming Automated Tutors",
booktitle="Artificial Intelligence in Education",
year="2018",
publisher="Springer International Publishing",
address="Cham",
pages="324--337",
isbn="978-3-319-93843-1"
}

@inproceedings{pldi18-automated,
author = {Gulwani, Sumit and Radi\v{c}ek, Ivan and Zuleger, Florian},
title = {Automated clustering and program repair for introductory programming assignments},
year = {2018},
isbn = {9781450356985},
publisher = {Association for Computing Machinery},
address = {New York, NY, USA},
url = {https://doi.org/10.1145/3192366.3192387},
doi = {10.1145/3192366.3192387},
booktitle = {Proceedings of the 39th ACM SIGPLAN Conference on Programming Language Design and Implementation},
pages = {465–480},
numpages = {16},
location = {Philadelphia, PA, USA},
series = {PLDI 2018}
}

@inproceedings{Frequence,
author = {Perer, Adam and Wang, Fei},
title = {Frequence: interactive mining and visualization of temporal frequent event sequences},
year = {2014},
isbn = {9781450321846},
publisher = {Association for Computing Machinery},
address = {New York, NY, USA},
url = {https://doi.org/10.1145/2557500.2557508},
doi = {10.1145/2557500.2557508},
booktitle = {Proceedings of the 19th International Conference on Intelligent User Interfaces},
pages = {153–162},
numpages = {10},
location = {Haifa, Israel},
series = {IUI '14}
}

@ARTICLE{ViSeq,
  author={Chen, Qing and Yue, Xuanwu and Plantaz, Xavier and Chen, Yuanzhe and Shi, Conglei and Pong, Ting-Chuen and Qu, Huamin},
  journal={IEEE Transactions on Visualization and Computer Graphics}, 
  title={ViSeq: Visual Analytics of Learning Sequence in Massive Open Online Courses}, 
  year={2020},
  volume={26},
  number={3},
  pages={1622-1636},
  doi={10.1109/TVCG.2018.2872961}}

@inproceedings{MatrixWave,
author = {Zhao, Jian and Liu, Zhicheng and Dontcheva, Mira and Hertzmann, Aaron and Wilson, Alan},
title = {MatrixWave: Visual Comparison of Event Sequence Data},
year = {2015},
isbn = {9781450331456},
publisher = {Association for Computing Machinery},
address = {New York, NY, USA},
url = {https://doi.org/10.1145/2702123.2702419},
doi = {10.1145/2702123.2702419},
booktitle = {Proceedings of the 33rd Annual ACM Conference on Human Factors in Computing Systems},
pages = {259–268},
numpages = {10},
location = {Seoul, Republic of Korea},
series = {CHI '15}
}

@inproceedings{Grotov2022,
author = {Grotov, Konstantin and Titov, Sergey and Sotnikov, Vladimir and Golubev, Yaroslav and Bryksin, Timofey},
title = {A large-scale comparison of Python code in Jupyter notebooks and scripts},
year = {2022},
isbn = {9781450393034},
publisher = {Association for Computing Machinery},
address = {New York, NY, USA},
url = {https://doi.org/10.1145/3524842.3528447},
doi = {10.1145/3524842.3528447},
booktitle = {Proceedings of the 19th International Conference on Mining Software Repositories},
pages = {353–364},
numpages = {12},
location = {Pittsburgh, Pennsylvania},
series = {MSR '22}
}

@misc{li2024unlockinginsightssemanticsearch,
      title={Unlocking Insights: Semantic Search in Jupyter Notebooks}, 
      author={Lan Li and Jinpeng Lv},
      year={2024},
      eprint={2402.13234},
      archivePrefix={arXiv},
      primaryClass={cs.IR},
      url={https://arxiv.org/abs/2402.13234}, 
}

@misc{siddik2025systematicliteraturereviewsoftware,
      title={A Systematic Literature Review of Software Engineering Research on Jupyter Notebook}, 
      author={Md Saeed Siddik and Hao Li and Cor-Paul Bezemer},
      year={2025},
      eprint={2504.16180},
      archivePrefix={arXiv},
      primaryClass={cs.SE},
      url={https://arxiv.org/abs/2504.16180}, 
}

@INPROCEEDINGS{11016521,
  author={Bajaj, Nikesh and Chiotis, Dimitrios and Moosaei, Reza and Smith, Jordan B. L. and Fan, Pengfei and Carrión, Jesús Requena},
  booktitle={2025 IEEE Global Engineering Education Conference (EDUCON)}, 
  title={A Tool for Detecting Similarities in Jupyter Notebooks Used as Assessment Reports}, 
  year={2025},
  pages={1-5},
  doi={10.1109/EDUCON62633.2025.11016521}
}

@unpublished{yang2025spark,
  author       = {Yang, Yinuo and Zhang, Ashley Ge and Oney, Steve and Wang, April Yi},
  title        = {SPARK: Real-Time Monitoring of Multi-Faceted Programming Exercises},
  note         = {To appear in IEEE Symposium on Visual Languages and Human-Centric Computing (VL/HCC)},
  year         = {2025},
  url          = {https://inonnno.github.io/Yinuo.me/static/media/spark_paper.e61cf1413a77a4df83e2.pdf}
}

@INPROCEEDINGS{1173155,
  author={Wattenberg, M.},
  booktitle={IEEE Symposium on Information Visualization, 2002. INFOVIS 2002.}, 
  title={Arc diagrams: visualizing structure in strings}, 
  year={2002},
  volume={},
  number={},
  pages={110-116},
  doi={10.1109/INFVIS.2002.1173155}}
\newpage
\appendix
\section{Finalized categories}
\label{appendix:classes}
\begin{table}[h]
\centering
\caption{Number of samples in each category after removing duplicates, sorted in descending order.}
\begin{tabular}{l r}
\hline
\textbf{Category} & \textbf{Count} \\
\hline
Data\_Transform       & 384220 \\
Visualization         & 373901 \\
Debug                 & 202776 \\
Model\_Train          & 186131 \\
Environment           & 101853 \\
Model\_Evaluation     &  95399 \\
Data\_Extraction      &  92557 \\
Feature\_Engineering  &  91792 \\
Other                 &  74235 \\
EDA                   &  65021 \\
Commented             &  48629 \\
Data\_Export          &  46322 \\
Hyperparam\_Tuning    &  37500 \\
\hline
\end{tabular}
\label{tab:category_counts}
\end{table}

\section{Complexity Feature Vector}
For each sequence \(A\), we compute the following features
\label{appendix:hmm}
\begin{enumerate}
    \item \textbf{Sequence Length}: Total sequence length (number of steps).
    \item \textbf{Unique States}: Count of distinct states present.
    \item \textbf{Diversity Ratio}: Unique states divided by length (a measure of coverage of state space).
    \item \textbf{Entropy}: Shannon entropy of the state distribution – higher entropy indicates more unpredictability in state occurrence.
    \item \textbf{State Changes}: Number of transitions where the state changes (i.e., how often the sequence switches state).
    \item \textbf{Change Ratio}: State changes normalized by sequence length.
    \item \textbf{Average Run Length}: Average length of consecutive repeats of the same state.
    \item \textbf{Maximum Run Length}: Maximum length of any run of identical states.
    \item \textbf{Run Length Variance}: Variability in the lengths of state-runs.
    \item \textbf{Pattern Complexity}: An approximation of Lempel–Ziv complexity measuring how many distinct subsequences appear (captures sequence novelty vs. repetitiveness).
    \item \textbf{Repetition Score}: Degree of repetitiveness, computed by finding the most repeated subsequence pattern (higher if a certain sub-pattern recurs frequently).
    \item \textbf{Transition Entropy}: Entropy of the transition matrix (i.e., uncertainty in what next state occurs, given the current state).
    \item \textbf{Self-Transition Ratio}: Fraction of transitions that stay in the same state.
    \item \textbf{Periodicity}: Presence of periodic patterns (e.g., how often elements repeat after a fixed interval).
    \item \textbf{Dominant State Ratio}: The relative frequency of the most common state, indicating if the sequence is dominated by one state or evenly spread.
\end{enumerate}

\section{Baseline Interface}
\begin{figure}[H]
    \centering
    \begin{subcaptionbox}{Notebook list in baseline interface}[0.48\textwidth]
        {\includegraphics[width=\linewidth]{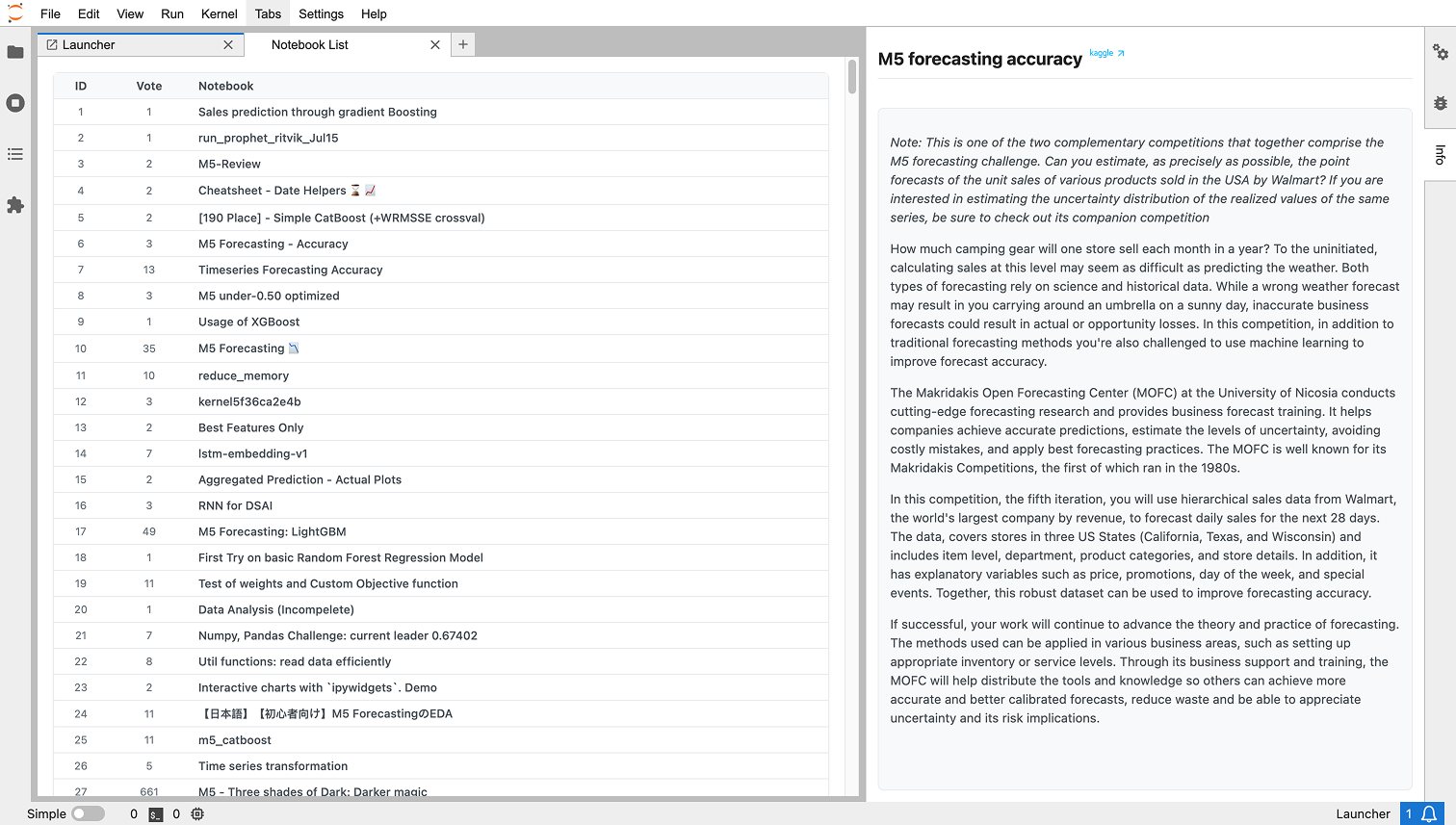}}
    \end{subcaptionbox}
    \hfill
    \begin{subcaptionbox}{Notebook content in baseline interface}[0.48\textwidth]
        {\includegraphics[width=\linewidth]{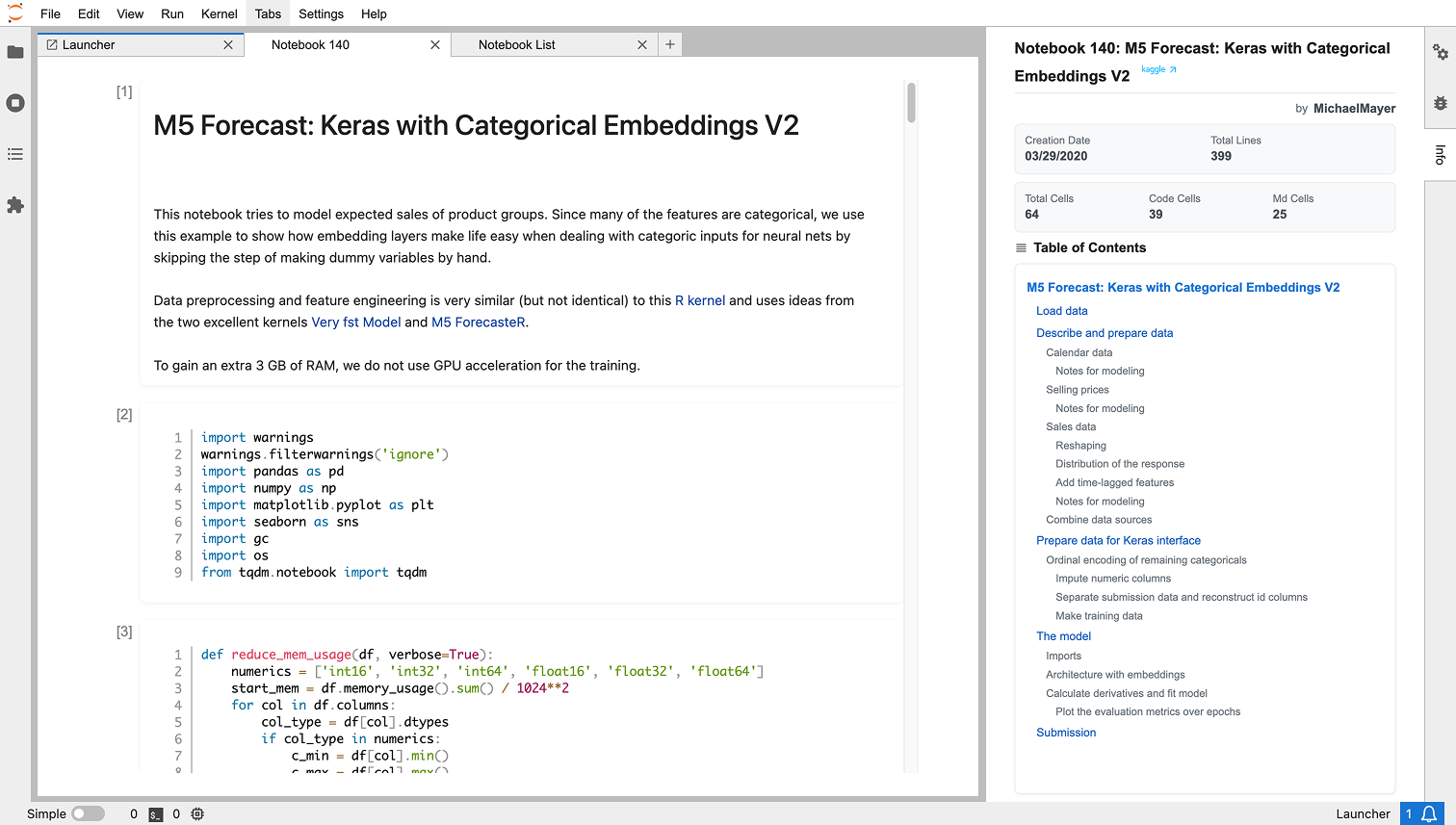}}
    \end{subcaptionbox}
    \caption{Baseline interface}
    \Description{Two screenshots （A, B) showing the baseline notebook interface. In image A, the central panel displays a list of notebooks for a Kaggle competition, while the right panel contains a textual description of the competition. In image B, the central panel shows the content of a selected notebook, including code and text cells, while the right panel presents basic notebook statistics and a table of contents.}
    \label{fig:baseline}
\end{figure}
\section{Split-screen Notebook View}
\begin{figure}[H]
    \centering
    \includegraphics[width=\linewidth]{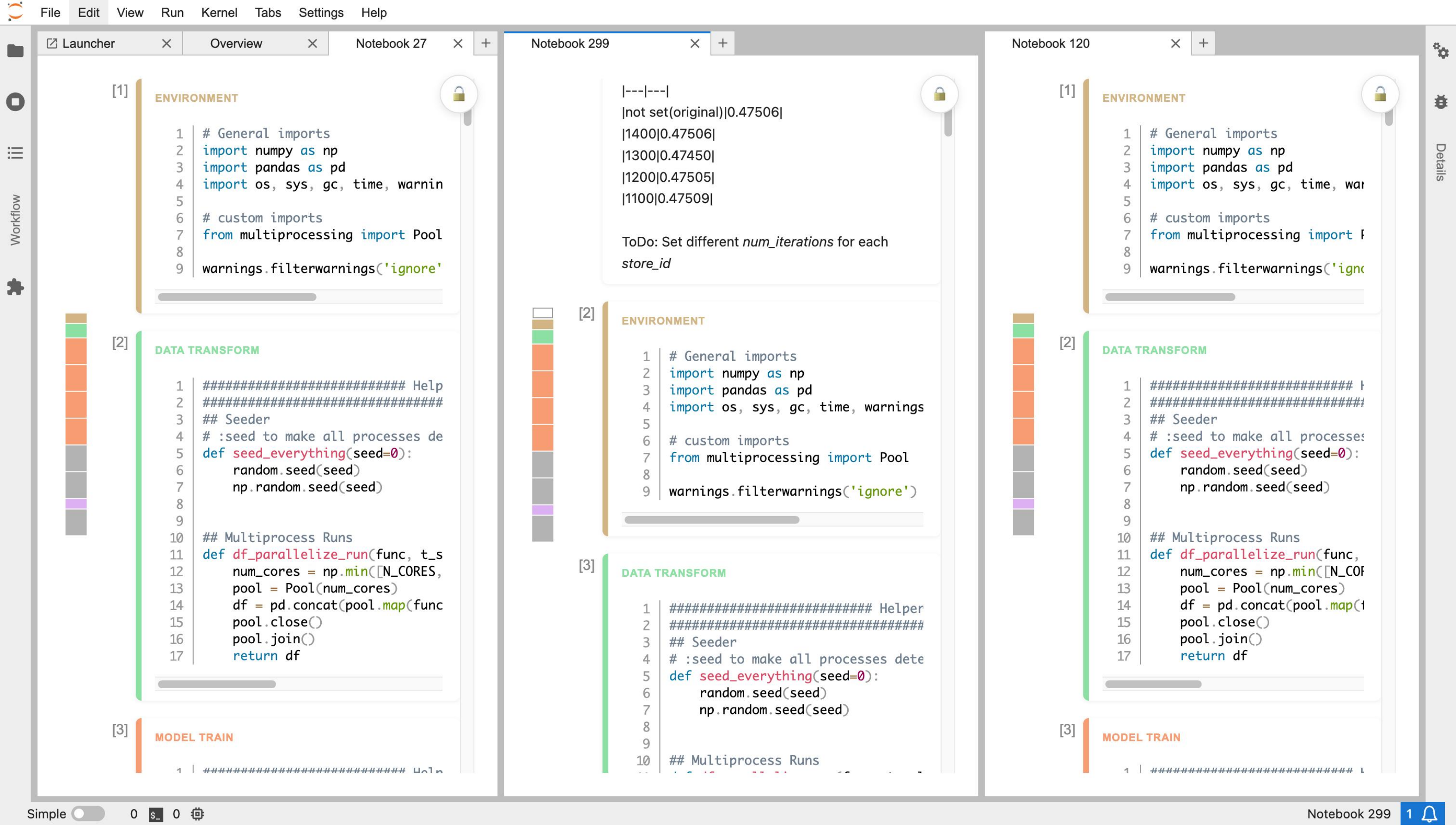}
    \caption{Split-screen Notebook View with synchronized scrolling, supporting close comparison of structurally similar notebooks.}
    \Description{Screenshot of the Split-screen Notebook View with three notebooks displayed side by side: Notebook 27 on the left, Notebook 299 in the middle, and Notebook 120 on the right. The middle notebook (299) is pinned as a fixed reference, while the left (27) and right (120) panes scroll in synchronization for comparison.}
    \label{fig:splitscreen}
\end{figure}

\section{Flowchart Figure}
\begin{figure}[H]
    \centering
    \begin{subcaptionbox}{Collection-level, hovering on a stage to highlight all corresponding cells in the matrix, clicking to fix the highlight.}[0.48\textwidth]
        {\includegraphics[width=\linewidth]{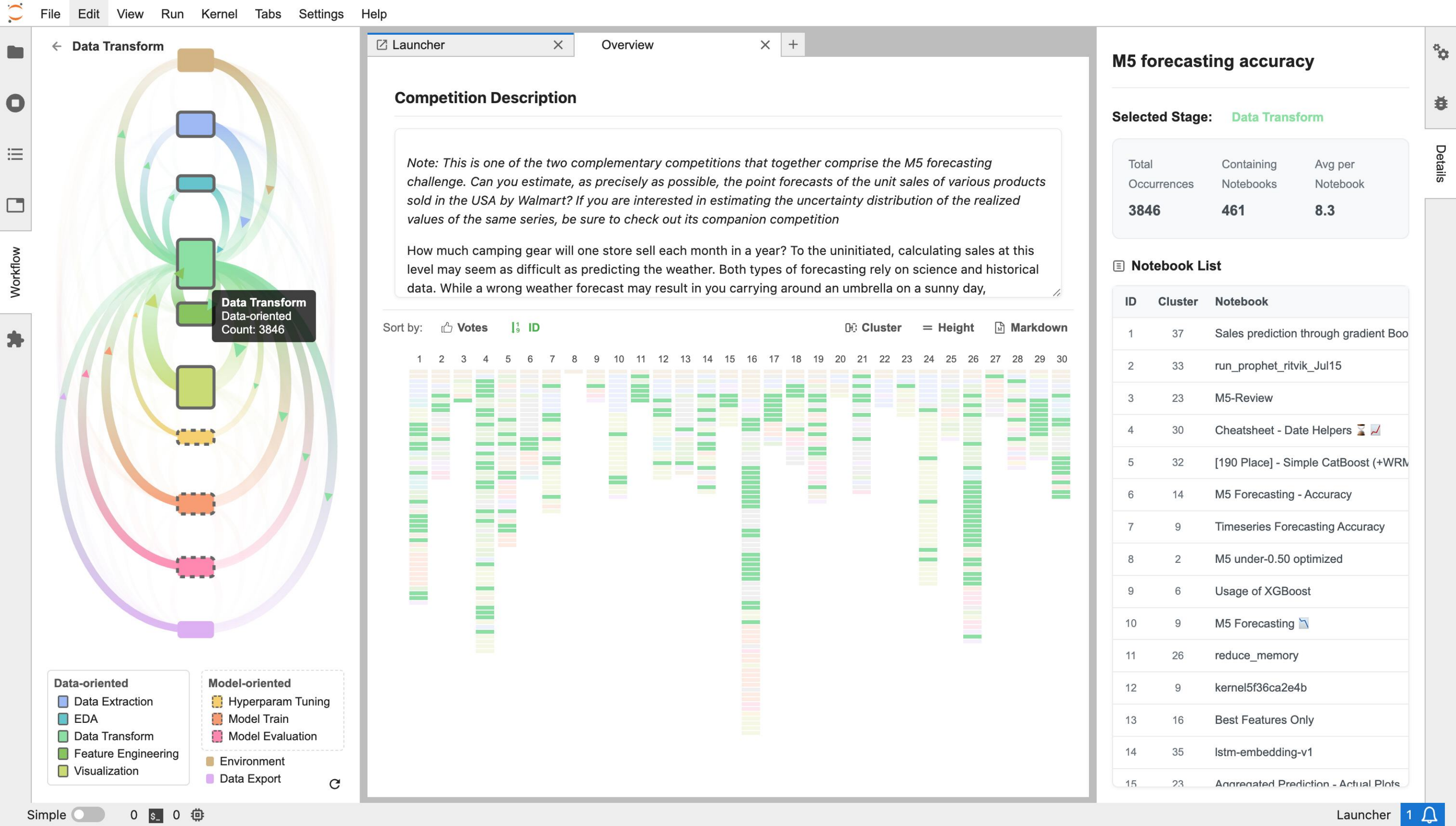}}
    \end{subcaptionbox}
    \hfill
    \begin{subcaptionbox}{Notebook-level, clicking a stage jumps to the first corresponding cell in the notebook, with a navigator at bottom to browse other cells of the same stage.}[0.48\textwidth]
        {\includegraphics[width=\linewidth]{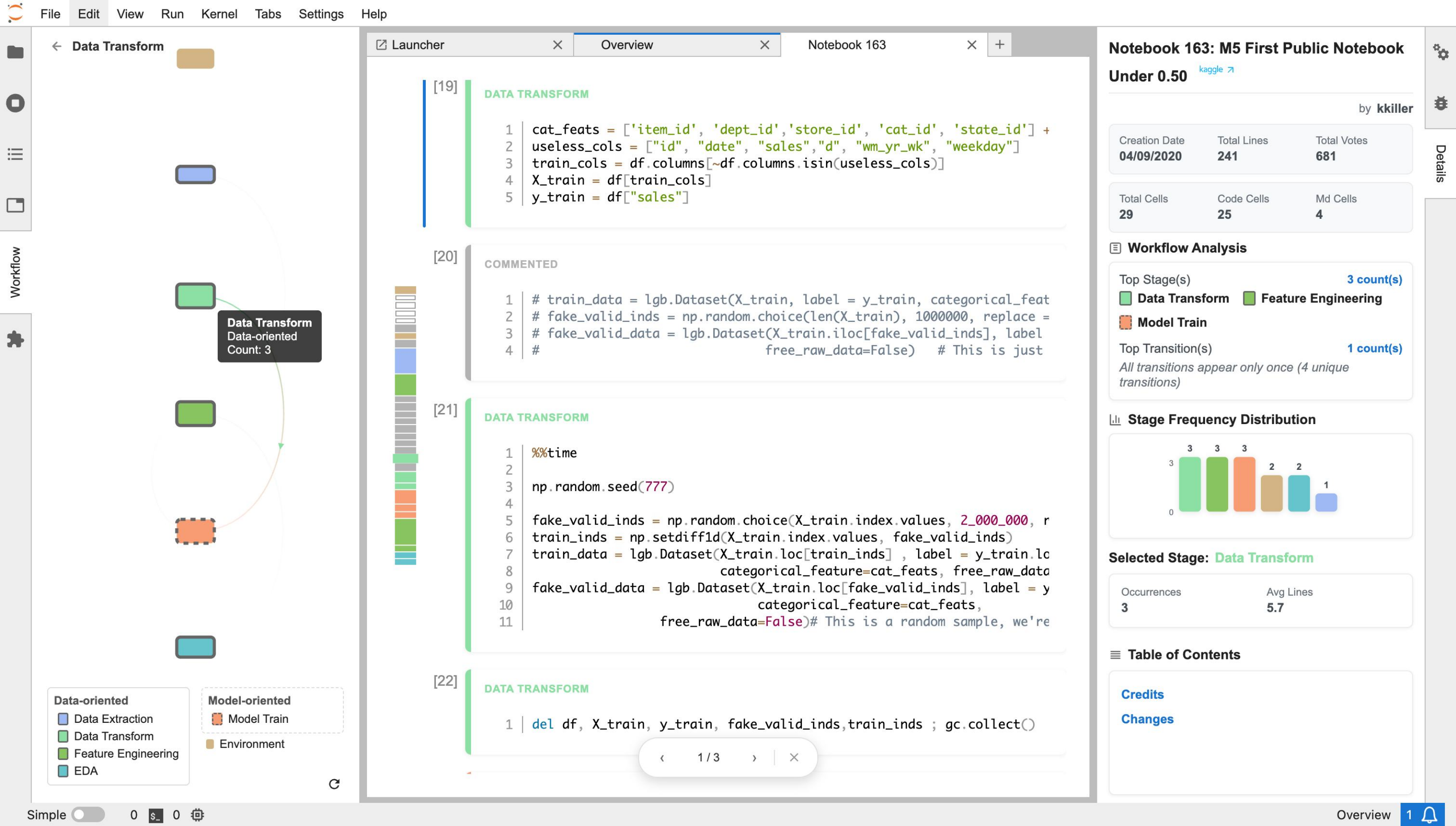}}
    \end{subcaptionbox}
    \caption{Arc-based workflow visualization for interactive stage selection and cross-view coordination.}
    \Description{Two screenshots labelled (A) and (B) show cross-view interactions in the \sys{} interface when the Data Transform stage is selected in the workflow view. Image (A) highlights all corresponding cells in the matrix view, while the detail view on the right shows the distribution of this stage across the entire collection. Image (B) shows the notebook view, where the system jumps to the first corresponding cell of the Data Transform stage, highlights it, and displays a navigation control for browsing other cells of the same stage.}
    \label{fig:crossview}
\end{figure}

\section{Comprehension Task}
\label{appendix:comprehension}

\textbf{Scenario.} You are reviewing solutions in a coding competition to compare how participants approached the same problem.

\textbf{Goals.} Analyze solution strategies at three levels: (1) an individual notebook, (2) a group of notebooks, and (3) the full collection.

\textbf{Part A: Understanding of the Competition}
\begin{enumerate}
    \item What is this competition about?
    \item What are the key objectives or requirements that need to be satisfied?
\end{enumerate}

\textbf{Part B: Single Notebook}

Please open notebook \textbf{N}.

\begin{enumerate}
    \item Locate all code cells relevant to the target stage (e.g., importing libraries, model training). Write down the ID numbers of these code cells, separated by commas.
    \item How confident are you in your answer?
    \item Compare notebook N with notebook M. Do they follow similar workflows?
    \item How confident are you in your answer?
    \item Briefly explain the reason for your comparison.
\end{enumerate}

\textbf{Part C: Several Notebooks}

Please review the following group of five notebooks (N1–N5).

\begin{enumerate}
    \item Skim through these notebooks. Which one has a noticeably different solution strategy from the others?
    \item How confident are you in your answer?
    \item Identify any nearly identical notebooks (similar workflows and similar cell counts). List all pairs or groups you find.
    \item How confident are you in your answer?
    \item Briefly explain why you believe those notebooks are similar.
\end{enumerate}

\textbf{Part D: All Notebooks}

Now consider the full set of notebooks in this competition. You do not need to inspect all notebooks exhaustively; provide your best estimation.

\begin{enumerate}
    \item Among the top 5 most-voted notebooks, which one contains the most markdown cells?
    \item How confident are you in your answer?
    \item Identify as many notebooks as possible where more than half of the code cells belong to the target stage (e.g., feature engineering, model training).
    \item How confident are you in your answer?
    \item Briefly explain your reasoning for selecting these notebooks.
\end{enumerate}

\section{Post-task Survey}
\label{appendix:survey}
\subsection{General Usability For Both Conditions}

\textbf{Part A: Interface Evaluation}

Please rate your agreement with each statement on a scale from 1 (Strongly disagree) to 7 (Strongly agree).
\begin{itemize}
    \item The interface is generally easy to use.
    \item The interface’s interaction is intuitive.
    \item It was easy for me to learn how to use the features in the interface.
    \item The interface helps me explore more alternative solutions effectively.
    \item The interface helps me identify similar solutions effectively.
    \item I am satisfied with my experience using the interface.
\end{itemize}

\textbf{Part B: Cognitive Load (NASA-TLX)}

Please rate each factor on a scale from 1 (Very low) to 7 (Very high).
\begin{itemize}
    \item How mentally demanding was the task?
    \item How hurried or rushed did you feel during the task?
    \item How hard did you have to work to accomplish your level of performance?
    \item How insecure, discouraged, irritated, stressed, or annoyed did you feel during the task?
\end{itemize}

\subsection{For \sys{}}
\label{appendix:notrix}
\textbf{Part A: Feature Usefulness}

Please rate the usefulness of each feature on a scale from 1 (Not useful at all) to 5 (Extremely useful).

\begin{itemize}
    \item ML stage classification (automatic labeling of code cells by workflow stage).
    \item Workflow visualization (visual representation of the notebook’s ML workflow).
    \item Side-by-side notebook comparison (visual and structural comparison of two notebooks).
    \item Similar pattern clustering (grouping notebooks with similar workflow patterns).
\end{itemize}

\textbf{Part B: Overall Perception}

Please rate your agreement with each statement on a scale from 1 (Strongly disagree) to 5 (Strongly agree).
\begin{itemize}
    \item The classification results produced by the system are reasonable.
    \item The clustering results produced by the system are reasonable.
    \item The clustering/grouping views revealed useful similarities and differences among notebooks.
    \item The interface helped me discover patterns in problem-solving strategies that I would not have otherwise noticed.
    \item The system supports exploring multiple notebooks well.
    \item The system supports understanding multiple notebooks.
    \item I would like to use a system like this in my work frequently.
\end{itemize}

\newpage
\onecolumn
\section{Classwise classification performance}
\label{appendix:clf}
\begin{table*}[ht]
\centering
\caption{Classification metrics by class label and overall averages on a holdout test set (in percentages)}
\begin{tabular}{lcccc}
\toprule
\textbf{Class} & \textbf{Precision (\%)} & \textbf{Recall (\%)} & \textbf{F1-Score (\%)} & \textbf{Accuracy (\%)} \\
\midrule
Environment          & 96.03 & 93.34 & 94.67 & 93.34 \\
Data\_Extraction     & 96.98 & 98.11 & 97.54 & 98.11 \\
Data\_Transform      & 87.38 & 83.02 & 85.15 & 83.02 \\
EDA                  & 95.24 & 97.98 & 96.59 & 97.98 \\
Feature\_Engineering & 90.27 & 89.87 & 90.07 & 89.87 \\
Model\_Train         & 84.51 & 83.93 & 84.22 & 83.93 \\
Hyperparam\_Tuning   & 89.75 & 90.74 & 90.24 & 90.74 \\
Model\_Evaluation    & 90.06 & 93.38 & 91.69 & 93.38 \\
Visualization        & 94.18 & 93.05 & 93.61 & 93.05 \\
Data\_Export         & 96.19 & 97.60 & 96.89 & 97.60 \\
Commented            & 99.22 & 99.89 & 99.55 & 99.89 \\
Debug                & 90.32 & 88.69 & 89.50 & 88.69 \\
Other                & 90.57 & 91.43 & 91.00 & 91.43 \\
\midrule
\textbf{Macro Average}    & 92.36 & 92.39 & 92.36 & 92.39 \\
\textbf{Weighted Average} & 92.36 & 92.39 & 92.36 & 92.39 \\
\bottomrule
\end{tabular}
\end{table*}

\section{ML stage label comparison}
\begin{table*}[ht!]
\centering
\renewcommand{\arraystretch}{1.2}
\caption{We showcase the refined 13 classes from Code4ML \cite{Code4ML} where we extracted \textit{Feature Engineering} and \textit{Commented} from the second-level label to the first-level label and eliminate the least common and underrepresented class \textit{Model Interpretation}. We compare this against categorisations from three different lines of work: ML lifecycle, DS workflows, and computational notebooks. We used all the refined labels in the classification but the grayed out stages were not used in the workflow visualization.}
\begin{tabular}{|>{\centering\arraybackslash}p{3.4cm}|>{\centering\arraybackslash}p{3.4cm}|>{\centering\arraybackslash}p{3.4cm}|>{\centering\arraybackslash}p{3.4cm}|}
\hline
\multirow{2}{*}{\makecell{\textbf{Refined labels} \\ \textbf{from Code4ML \cite{Code4ML}}}} & \multicolumn{3}{c|}{\textbf{Corresponding labels identified from the literature}} \\ 
\cline{2-4}
& \textbf{ML lifecycle \cite{sigmod23-ml-liftcycle, csur21-ml-lifecycle, dis24-ds-hai}} & \textbf{DS Workflow \cite{ramasamy2023visualising, ese23-workflow, aggarwal2019can}} & \textbf{Computational Notebooks \cite{cscw19-hai-ds, zhang2022coral}} \\ 
\hline
Environment & - & Helper Functions & - \\ \hline
Data Extraction & Data Collection and Selection & Load Data & Data Acquisition \\ \hline
Data Transform & Data Cleaning \& Labeling & Data Preprocessing & Data Cleaning \& Labeling \\ \hline
Exploratory Data Analysis & - & Data Exploration & - \\ \hline
Feature Engineering & Feature Engineering and Selection & - & Feature Engineering \\ \hline
Visualization & - & Result Visualization & - \\ \hline
Model Train & Model Training & Modeling & - \\ \hline
Hyperparameter Tuning & Model Optimization & - & Hyperparameter Optimization \\ \hline
Model Evaluation & Model Evaluation & Evaluation & Model Validation \\ \hline
Data Export & - & Save Results & - \\ \hline
\textcolor{gray}{Debug} & - & - & - \\ \hline
\textcolor{gray}{Commented} & - & Comment Only & - \\ \hline
\textcolor{gray}{Other} & - & - & - \\ \hline
\end{tabular}
\label{tab:labels}
\end{table*}

\newpage
\section{Selected Competitions}
\begin{table*}[ht]
\centering
\caption{Selected competitions, data downloaded 22nd June 2025. The first two competitions were used in the comprehension task and the last competition were used in the practice task for warm up.}
\begin{tabular}{ccccc}
\toprule
\textbf{ID} & \textbf{Title} & \textbf{Enabled Date} & \textbf{Submissions} & \textbf{Notebooks} \\
\midrule
18599 & \href{https://www.kaggle.com/competitions/m5-forecasting-accuracy}{M5 forecasting accuracy} & 2020-03-03 & 109451 & 521 \\
50160 & \href{https://www.kaggle.com/competitions/home-credit-credit-risk-model-stability}{Home Credit - Credit Risk Model Stability} & 2024-02-05 & 98711 & 420 \\
35332 & \href{https://www.kaggle.com/competitions/amex-default-prediction}{American Express} & 2022-05-25 & 108474 & 602 \\
\bottomrule
\end{tabular}
\label{tab:competitions}
\end{table*}

\clearpage
\section{Cross-Level Navigation Across Task Scopes}
\label{appendix:navigation}
\begin{figure}[htbp]
    \centering
    \includegraphics[width=\linewidth,height=0.78\textheight,keepaspectratio]{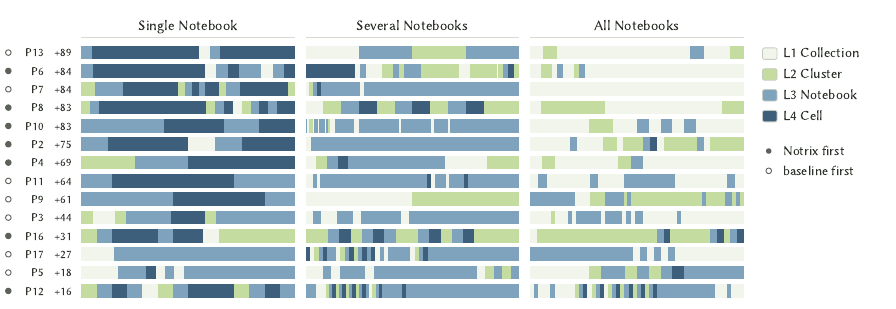}
    \caption{User navigation across three task scopes (\sys{} condition, $N = 14$). The number beside each participant is $p_{\text{All}} - p_{\text{Single}}$ in percentage points, where $p$ is the share of that participant's interactions in a task that fell at the collection and cluster levels (L1 and L2): how much more of their activity sat in the two broader views when the question widened from one notebook to all notebooks. Rows are sorted by it. All 14 are positive (median 82\% in All Notebooks against 16\% in Single Notebook; Wilcoxon signed-rank, $W = 0$, $p < .001$). Filled dots mark participants who used \sys{} before the baseline; tasks appear in the order they were administered.}
    \Description{User navigation across three task scopes (\sys{} condition, $N = 14$). The number beside each participant is $p_{\text{All}} - p_{\text{Single}}$ in percentage points, where $p$ is the share of that participant's interactions in a task that fell at the collection and cluster levels (L1 and L2): how much more of their activity sat in the two broader views when the question widened from one notebook to all notebooks. Rows are sorted by it. All 14 are positive (median 82\% in All Notebooks against 16\% in Single Notebook; Wilcoxon signed-rank, $W = 0$, $p < .001$). Filled dots mark participants who used \sys{} before the baseline; tasks appear in the order they were administered.}
    \label{fig:time}
\end{figure}

\end{document}